\documentclass[onecolumn]{aastex63}
\usepackage[utf8]{inputenc}
\usepackage{amsmath,amsfonts,amssymb}
\usepackage{graphicx,subfigure,multirow}
\usepackage{epsfig,color}
\usepackage{enumitem}
\usepackage{natbib}
\usepackage{hyperref}
\usepackage{verbatim}
\usepackage[bottom]{footmisc}

\usepackage{slashbox}

\usepackage{pbox}

\newcommand{\vek}[1]{\vec{#1}}

\begin{document}

\title{Bayesian Inference of Dense Matter Equation of State within Relativistic Mean Field Models using Astrophysical Measurements}


\author[0000-0002-8410-520X]{Silvia Traversi}
\affiliation{Dipartimento di Fisica e Scienze della Terra,  Università di Ferrara, Via Saragat 1, 44122 Ferrara, Italy}
\affiliation{INFN Sezione di Ferrara, Via Saragat 1, 44122 Ferrara, Italy}

\author[0000-0001-6592-6590]{Prasanta Char}
\affiliation{INFN Sezione di Ferrara, Via Saragat 1, 44122 Ferrara, Italy}

\author[0000-0003-3250-1398]{Giuseppe Pagliara}
\affiliation{Dipartimento di Fisica e Scienze della Terra,  Università di Ferrara, Via Saragat 1, 44122 Ferrara, Italy}
\affiliation{INFN Sezione di Ferrara, Via Saragat 1, 44122 Ferrara, Italy} 


\begin{abstract}
We present a Bayesian analysis to constrain the equation of state of dense nucleonic matter by exploiting the available data from symmetric nuclear matter at saturation and from observations of compact X-ray sources and from the gravitational wave event GW170817. For the first time, such analysis is performed by using a class of models, the relativistic mean field models, which allow to consistently construct an equation of state in a wide range of densities, isospin asymmetries and temperatures. The selected class of models contains five nuclear physics empirical parameters at saturation for which we construct the joint posterior distributions. By exploring different types of priors, we find that the equations of state with the largest evidence are the ones featuring a strong reduction of the effective mass of the nucleons in dense matter which can be interpreted as an indication of a phase transition to a chiral symmetry restored phase. Those equations of state in turn predict $R_{1.4} \sim 12$ km. Finally, we present a preliminary investigation on the effect of including $\Lambda$ hyperons showing that they appear in stars more massive than about $1.6 M_{\odot}$ and lead to radii larger than about $R_{1.4} \sim 14$ km. Within the model here explored, the formation of such particles provide a poor agreement with the constraints from GW170817. 
\end{abstract}


\section{Introduction}
A neutron star (NS), born in the aftermath of a core-collapse supernova explosion, can sustain densities above a few times the nuclear saturation density in its interior  \citep{Glendenning:1997wn}. The composition and the properties of matter are largely unknown at such densities. The data available from laboratory experiments and \textit{ab initio} calculations provide the descriptions of the equation of state (EOS) of symmetric nuclear matter at the nuclear saturation density. Then, the EOS is extrapolated using available theoretical models to describe the properties of matter at supranuclear densities and high isospin asymmetries, but remains largely model dependent. Fortunately, we have a plethora of astronomical observations that can help to constrain the EOS for NS matter \citep[see][for a review]{Lattimer:2006xb, Oertel:2016bki}. Since one can have a unique map from the pressure-energy density relation to the mass-radius diagram by solving the Tolman-Oppenheimer-Volkoff (TOV) equations, the measurements of masses and radii can provide valuable insights on the EOS at that density regime. Presently, we have precise observations of masses for several massive NSs \citep{Demorest:2010bx,Antoniadis:2013pzd,Fonseca:2016tux,Arzoumanian:2017puf,Cromartie:2019kug}. The observations of pulsars heavier than $2 M_\odot$ have put a very strong lower limit on the maximum mass and have already ruled out many soft EOSs. There exist a few independent observations for NS radii as well, from the thermonuclear bursts emitted by accreting NSs and from the thermal emission of low-mass X-ray binaries in quiescence (qLMXB)  \citep{Guillot:2013wu,Ozel:2015fia,Ozel:2016oaf,Nattila:2017wtj}. Most recently, the NICER collaboration has also reported a very accurate joint measurement of the mass and the radius of the millisecond pulsar PSR J0030+0451 \citep{Riley:2019yda,Miller:2019cac}. In 2017, LIGO-VIRGO collaboration (LVC) has reported the first ever detection of gravitational waves (GW) from a binary neutron star (BNS) merger event, GW170817 which has provided an estimate of the combined tidal deformability for the components of the binary \citep{TheLIGOScientific:2017qsa, Abbott:2018exr}.  The tidal deformability and the radii measurements tend to prefer more compact stars, therefore several stiff EOSs have been ruled out. In this regard, future observations from NICER and other X-ray missions, such as eXTP \citep{Watts:2018iom} will be able to provide simultaneous measurements of mass and radius with better accuracy thus allowing to obtain even stronger constraints on the EOS.

In recent years, since the seminal work of \citet{Steiner:2010fz}, a substantial progress has been made to provide statistical inference on the EOS models in the light of the available astrophysical data and the experimental laboratory data. The Bayesian method is appropriate in this context because it allows to naturally include the a priori knowledge on the low density and symmetric EOS and to explore, in a controlled way, the wide space of parameters which determine the neutron star EOS. The new paradigm, proposed in  \citet{Steiner:2010fz}, is to match the low density EOS (which is constrained by theoretical and experimental nuclear physics) with parametrized high density EOSs. A very simple choice is to use piecewise polytropic EOSs, since with a few parameters (for instance the adiabatic indices and the densities of matching) one can reproduce within a small error most of the proposed theoretical calculations of EOSs available in the literature \citep{Read:2008iy}. Interestingly, these parametrizations can also be constrained in the very high density regime for which perturbative QCD calculations are available \citep{Kurkela:2014vha}. 
A second possibility, is to use the so-called spectral representation of the EOS which is based on a series expansion of the adiabatic index 
\citep{Lindblom:2010bb,Fasano:2019zwm}. Also in this case, few terms in the series are needed in order to capture the high variability of the theoretical EOSs.
Recently, a new scheme has been adopted in \cite{Capano:2019eae}, in which 
$\chi$EFT results are adopted up to twice saturation density and a sampling of the speed of sound values is performed for larger densities.
A nonparametric inference approach has been followed instead in \citet{Essick:2019ldf}, the major advantage being the possibility of generating directly the EOS without the need of parameters such as adiabatic indices and/or speed of sound at specific values of the density. Finally, machine learning techniques are also started being used in \citet{Fujimoto:2017cdo,Fujimoto:2019hxv,Ferreira:2019bny}.

Another powerful method proposed in the literature is the metamodelling of \citet{Margueron:2017eqc,Margueron:2017lup} where the EOS is built up from a Taylor expansion of the energy per baryon from saturation to high density and high isospin asymmetries. Clearly, those methods have advantages and drawbacks: the metamodelling offers the possibility to fully exploit the experimental results on symmetry energy and its density dependence and also to compute the composition of beta stable matter, in particular the proton fraction, which in turn allows to constrain the EOS not only via masses, radii and tidal deformabilities but also with neutron stars cooling data. This aspect is completely neglected within piecewise polytropic and spectral approaches in which only the structure of neutron stars can be computed. Also, these parametrizations do not contain any aspect of the physics of the symmetry energy. On the other hand, the scheme proposed in \citet{Margueron:2017eqc,Margueron:2017lup} deals with non-relativistic nucleons and therefore it can not mimic (in the present version) either the appearance of new degrees of freedom (such as hyperons) or a possible 
phase transition to new phases such as quark matter \citep[see][for instance]{Alvarez-Castillo:2016oln}. Phase transitions can, on the other hand, be included e.g. in piecewise polytropic methods. Indeed, some sources can be better modelled with strong phase transition as found in \citep{Steiner:2017vmg}.

In this paper, we propose an alternative scheme for a statistical inference of the EOS based on a class of Relativistic Mean Field  (RMF) models in which the interaction between baryons is mediated by the exchange of scalar and vector mesons.
These models, which are extensions of the Walecka model, 
have been largely adopted for the calculations of finite nuclei and for the modeling of the EOS within a wide range of densities, temperatures and chemical compositions.
Many supernova and merger simulations have indeed used those kind of EOSs
such as e.g. the SFHo \citep{Steiner:2012rk}, the BHB$\Lambda\phi$ \citep{Banik:2014qja} and the TM1 models \citep{Shen:1998gq}.
The advantages of such models are that they can encode the constraints from symmetry energy and finite nuclei, they can easily include hyperons and other possible baryons and also the chemical composition and its impact on the cooling can be studied.
In particular, one can include the (few) information we have from the experimental data on hypernuclei. These models have also been confronted recently with the GW data \citep{Malik:2018zcf,Nandi:2018ami, Lourenco:2018dvh}. Another aspect concerns phase transitions: the mean field equation for the $\sigma$ field is a non-linear equation which could mimic the occurrence of a phase transition if one uses the nucleon effective mass as an order parameter.
As we will discuss later, a rapid drop of the effective mass as a function of the baryon density could be interpreted as suggesting partial restoration of chiral symmetry and thus a likely phase transition to quark matter.
Another interesting aspect is that, once the zero temperature EOS has been constrained by neutron stars observations, one can extend it to finite temperatures and test it also for studies of transient phenomena such as merger and supernova events.

Concerning the astrophysical data, we use a larger data set in our Bayesian analysis with respect to previous works. In particular, we have included the data for the six  thermonuclear bursters and five qLMXBs from \cite{Ozel:2015fia}, one X-ray source of \cite{Nattila:2017wtj} and the two components of GW170817 \citep{Abbott:2018exr}. We also include the recent NICER observations of PSR J0030+0451.

The paper is organized as follows: in Section \ref{EOS}, we introduce the RMF parametrization we adopt for performing the Bayesian analysis. In Section \ref{bayes}, after a brief review on the method, we will present the different priors on the EOS and the astrophysical data which will be adopted for constructing the likelihood. In Section \ref{result} we will present our results and in Section \ref{discussion} the discussion and the conclusions. In the appendix, we provide further details of our calculations.


\section{Equation of State}\label{EOS}
The class of RMF models we adopt in this paper is the non-linear Walecka model proposed in the seminal paper of \citet{Boguta:1977xi,Glendenning:1991es} which is characterized by the presence of a cubic and a quartic self-interaction terms for the scalar field $\sigma$. Those two terms have been added to the Walecka model in order to
keep the values of incompressibility and effective mass at saturation under control \citep{Boguta:1977xi}.
The Lagrangian of the model is given by, 

\begin{eqnarray}\label{lag}
{\cal L}_B &=& \sum_{B} \bar\Psi_{B}\left(i\gamma_\mu{\partial^\mu} - m_B
+ g_{\sigma B} \sigma - g_{\omega B} \gamma_\mu \omega^\mu
- g_{\rho B}
\gamma_\mu{\mbox{\boldmath $\tau$}} \cdot
{\mbox{\boldmath $\rho$}}^\mu \right)\Psi_B\nonumber\\
&& + \frac{1}{2} \left(\partial_\mu \sigma\partial^\mu \sigma
- m_\sigma^2 \sigma^2 \right) - \frac{1}{3}b m_n \left(g_\sigma \sigma\right)^3 
- \frac{1}{4}c  \left(g_\sigma \sigma\right)^4 \nonumber\\
&& -\frac{1}{4} \omega_{\mu\nu}\omega^{\mu\nu}
+\frac{1}{2}m_\omega^2 \omega_\mu \omega^\mu
- \frac{1}{4}{\mbox {\boldmath $\rho$}}_{\mu\nu} \cdot
{\mbox {\boldmath $\rho$}}^{\mu\nu}
- \frac{1}{2}m_\rho^2 {\mbox {\boldmath $\rho$}}_\mu \cdot
{\mbox {\boldmath $\rho$}}^\mu ~,
\end{eqnarray}

where, the $\sigma$, $\omega$ and  $\rho$ represent the scalar, vector and vector-isovector mesons respectively. 



The model contains five unknown parameters: $g_\sigma/m_\sigma$, $g_\omega/m_\omega$, $g_\rho/m_\rho$, $b$, $c$. These quantities are related algebraically to five saturation properties of symmetric nuclear matter i.e. the binding energy per nucleon ($E_0$), the saturation density ($n_0$), the symmetry energy ($S$), the incompressibility ($K$) and the effective mass of nucleon ($m^*$) \citep{Glendenning:1997wn}. Hence, we can effectively parametrize our EOS in terms of the aforementioned empirical nuclear physics parameters. 
We do not report here the expressions for the mean field equations 
and for the thermodynamical quantities that can be found in many references, 
see e.g. \cite{Glendenning:1997wn}.
The model can be easily extended to include hyperons.
We consider here only the case of the $\Lambda$ hyperon whose experimental value for the potential depth is fixed to $-28$MeV as in \citet{Glendenning:1991es}. The coupling with the $\omega$ meson is set to its SU(3) symmetry value i.e. $2/3 g_\omega/m_\omega$  \citep{Weissenborn:2011ut} whereas the coupling with the $\sigma$ meson is directly derived from the potential depth.
In this paper, for the Bayesian analysis with the inclusion of hyperons,
we do not vary the potential depth nor the coupling with the $\omega$ meson beyond flavor symmetry, as in \citet{Weissenborn:2011ut}.
Thus we just want to study the effect on the posterior of switching on 
the $\Lambda$'s. In a forthcoming paper, we will investigate also the effects of changing those couplings and of including all the baryons of the octet.

A major drawback of this model is that it cannot include other 
saturation properties such as the slope of the symmetry energy $L$ or higher order derivatives (curvature $K_{\mathrm{sym}}$ and skewness $J_{\mathrm{sym}}$) which are shown to be relevant in \cite{Xie:2019sqb} in the density range up to $2.5 n_0$. 
In particular, in our parametrization $L$ is a derived quantity and is correlated with $S$. We anticipate that in some of our results it will come out that $L$ is larger than the typical values proposed in the literature i.e. $40$ MeV $\lesssim L \lesssim 60$ MeV,  \citep[see][for details]{Xie:2019sqb,Lattimer:2012xj}. The tidal deformability of GW170817 also seems to suggest even smaller values, with the range of $L$ extending down to $9$ MeV \citep{Raithel_2019} thus pointing to a significant tension with laboratoty data on neutron skin thickness  \citep[see][for a discussion]{Fattoyev:2017jql}. 
In principle, to include such quantities one would need to introduce other interaction terms (including meson mixing terms as in \cite{Steiner:2012rk}) or density dependent couplings as in \cite{Typel:1999yq} with the cost of loosing the direct analytical expressions connecting the parameters of the model and the saturation properties. We postpone such an extension of our work for a future paper. Another drawback is connected with the sign of the quartic coupling $c$. In several accepted parametrizations it is negative thus making the energy unbounded from below for a large value of the $\sigma$ field. A common viewpoint concerning this issue is to consider the model as an effective  model which cannot be extrapolated to arbitrarily high densities and to check for well-behaved EOS  (causality limit and absence of mechanical and chemical instabilities) in the density regime of compact stars. Again, also this problem could be addressed in future studies by using hadronic chiral models, such as in \citet{Bonanno:2008tt}.

\section{Bayesian Analysis}\label{bayes}
In Bayesian parameter estimation, the probability distribution of a set of model parameters $\vek{\theta}$, called the posterior density function (PDF) given a data ($D$) for some model ($M$) is inferred by the Bayes' theorem,

\begin{equation} \label{E:posterior}
    P(\vek{\theta}|D,M) = \frac{P(D|\vek{\theta},M)P(\vek{\theta}|M)}{P(D|M)},
\end{equation}
where $P(\vek{\theta}|M)$  is the prior probability of the parameter set $\vek{\theta}$. This is updated by the experimental data through the likelihood function, $P(D|\vek{\theta},M)$. The denominator, $P(D|M)$ is known as evidence for the model $M$ which is a constant for the given data $D$. Since it does not depend on the model parameters, the evidence can be treated as a normalization factor for the posterior. In this work, we mainly follow the Bayesian methodology developed by \cite{Steiner:2010fz,Ozel:2015fia,Raithel:2017ity} which is described in the following.

Our parameter set $\vek{\theta}$ consists of the empirical parameters: $\{m^*, K, n_0, S, E_0\}$. In order to calculate the posterior over $P(\vek{\theta})$ using the mass-radius distributions $P_i(M,R)$ for the $N=15$ sources which we have used in our analysis (see next Sec.), we can write,
\begin{equation} \label{E:poste_para}
    P(m^*, K, n_0, S, E_0|data) = C P(data|m^*, K, n_0, S, E_0) \times  P(m^*) P(K) P(n_0) P(S) P(E_0) ,
\end{equation}
where $C$ is the normalization constant; $P(m^*)$, $P(K)$, $P(n_0)$, $P(S)$, and $P(E_0)$ are the priors over the empirical parameters and
\begin{equation}
    P(data|m^*, K, n_0, S, E_0) = \prod_{i=1}^{N} P_i(M_i,R_i|m^*, K, n_0, S, E_0),
\end{equation}
is the likelihood of generating $N$ mass-radius observations given a particular set of empirical parameters. To compute the probability of the realization of $(M, R)$ for a particular source given an EOS, we follow the procedure suggested by \cite{Raithel:2017ity}. We take a set of parameters and calculate the couplings using the algebraic relations. Then, we solve the mean field equations to construct the EOS. Using the EOS, we solve the TOV equations and build a mass-radius curve up to the maximum mass which corresponds to the last stable point of the curve. After that, we compute the probability of each configuration of the curve using the M-R distribution of the source. Finally, we assign to the parameter set the maximum probability obtained for the configurations as,

\begin{equation}
    P_i(M_i,R_i|m^*, K, n_0, S, E_0) = P_{\text{max}}(M_i,R_i|m^*, K, n_0, S, E_0, \rho_c),
\end{equation}

where, the mass-radius curve for a given EOS is parametrized by the central energy density ($\rho_c$) of the star.  We use the Markov-Chain Monte Carlo (MCMC) simulations to populate the posterior distribution of Equation (\ref{E:poste_para}) using the python \textsf{emcee} package with stretch-move algorithm \citep{ForemanMackey:2012ig}. Next, we calculate the evidence to compare between different models by performing a Monte Carlo integration over the posterior. We also compute the Bayesian information criterion using the standard definition: BIC $\equiv -2 \ln {\cal L}_{max} + k \ln N$, where ${\cal L}_{max}$ is the maximum likelihood, $k$ the number of parameters, and  $N$ the number of data points.

\subsection{Conditions on the priors}\label{prior}
The choice of priors plays an important role to interpret the data within the domain of a model. Ambiguities in the priors can alter the shape of the posteriors and  consequently the model predictions \citep{Steiner:2015aea}. Therefore, we need to study the effects of different prior types and prior ranges. These priors essentially encompass our assumptions for the model. Hence, they can be treated as individual models by themselves and we compute the Bayes factors between every pair of them for a quantitative comparison. We categorize our priors into two different classes: informed and agnostic. For the first class, we choose our priors based on the available  constraints from laboratory experiments on the nuclear empirical parameters. For the agnostic priors, we relax those constraints and allow the parameters to be determined mainly from the requirements of the astrophysical data. We implement both Gaussian and  uniform priors to investigate their effects on the parameter estimation. In particular, we take five different priors motivated by several different studies in literature \citep{Lattimer:2012xj,Dutra:2014qga,Oertel:2016bki,Margueron:2017eqc}. 

\begin{description}[itemsep=0pt]
    \item[Baseline] The prior ranges are inspired from the results of laboratory experiments. We take the ranges of $m^*$, $K$ from \cite{Glendenning:1997wn} and $S$ from \cite{Lattimer:2012xj}. Additionally, we choose a sensible range for both $n_0$ and $E_0$. 
    \item[Marg\_unif] Following the Table 11 of \cite{Margueron:2017eqc} concerning the RMF models, we define a uniform prior with the minimum and maximum value of the parameters of our interest. Those models were chosen among a wider class of RMF models for their ability to provide sensible results for a large number of nuclear properties.
    
    \item[Marg\_Gauss]  From the aforementioned table of \cite{Margueron:2017eqc}, we take the mean and standard deviation of our parameters to define a Gaussian prior.
    
    \item[Wide\_unif]  We consider the type 2 EOSs of Table VII of \cite{Dutra:2014qga} and define a uniform prior over the maximum and minimum values of the parameters. We consider only the type 2 EOSs because those correspond to the RMF model with the same scalar self-interactions as in our case. We remark that not all the models of this table are compatible with modern nuclear physics constraints particularly concerning the symmetry energy. 
    
    \item[Wide\_Gauss]  We consider again the type 2 EOSs of Table VII of \cite{Dutra:2014qga} and parametrize the uncertainties with Gaussian distributions. In this case, the standard deviations are somewhat larger than the Marg\_Gauss.
\end{description}

The corresponding ranges of the empirical parameters for the priors mentioned above are listed in Table \ref{tab:prior}. 

\begin{table}[h]

    \begin{tabular}{|c|c|c c c c c|}
    \hline
    \hline 
    Priors & Range & $m^*$ & $K$ (MeV) & $n_0$ ($fm^{-3}$) & $S$ (MeV) & $B/A$ (MeV) \\
    \hline 
    Baseline  & Min & $0.7$ & $200$ & $0.14$ & $28$ & $-16.5$ \\
    \cline{2-7}
            & Max & $0.8$ & $300$ & $0.16$ & $35$ & $-16.0$ \\
    \hline
    Marg\_unif  & Min & $0.64$ & $219$ & $0.145$ & $31.19$ & $-16.35$ \\
    \cline{2-7}
            & Max & $0.71$ & $355$ & $0.153$ & $38.71$ & $-16.12$ \\
    \hline 
    Marg\_Gauss  & Mean & $0.67$ & $268$ & $0.1494$ & $35.11$ & $-16.24$ \\
    \cline{2-7}
            & $\sigma$ & $0.02$ & $34$ & $0.0025$ & $2.63$ & $0.06$ \\
    \hline 
    Wide\_unif  & Min & $0.55$ & $172.23$ & $0.145$ & $17.38$ & $-17.03$ \\
    \cline{2-7}
            & Max & $0.8$ & $421.02$ & $0.173$ & $50.0$ & $-13.78$  \\
    \hline 
    Wide\_Gauss  & Mean & $0.708$ & $245.29$ & $0.152$ & $34.11$ & $-16.17$  \\
    \cline{2-7}
            & $\sigma$ & $0.079$ & $39.30$ & $0.004$ & $4.42$ & $0.36$ \\
    \hline    
    \end{tabular}
    \caption{List of the priors that we employ in this work.}
    
    \label{tab:prior}
\end{table}

Additionally, we impose the following physical constraints and observational requirements on the EOSs that we construct using the priors. 
\begin{enumerate}[label=\roman*., itemsep=0pt]
\item The EOSs should be mechanically stable.
\item The EOSs must remain causal for the range of densities of our interest.
\item The maximum stable mass produced by each EOS must be compatible with the observations. In this work we impose the condition that it must exceed $2 M_\odot$, as in \citet{Guven:2020dok}.
\end{enumerate}

\subsection{Observational Data}\label{sources}
In this section, we briefly describe the fifteen sources used in our work. In \cite{Ozel:2015fia}, measurements of two different types of sources are reported: the thermonuclear bursters and the qLMXBs. \footnote{The M-R distributions of the sources of \cite{Ozel:2015fia} are available at \href{http://xtreme.as.arizona.edu/neutronstars/}{http://xtreme.as.arizona.edu/neutronstars/}.} The burst sources are 4U 1820--30, SAX J1748.9--2021, EXO 1745--248, KS 1731--260, 4U 1724--207, and 4U 1608--52. The mass-radius posteriors for these sources are calculated using their apparent angular sizes, touchdown fluxes, and distances. The qLMXBs are M13, M30, NGC 6304, NGC 6397, and $\omega$ Cen. In quiescence, the heat accumulated in the crusts of the stars during accretion is radiated through a light element atmosphere. The mass-radius posteriors of the qLMXBs are inferred from the spectral analysis of the thermal emission.

Another source of information comes from the X-ray burst cooling tail spectra of the  NS in 4U 1702--429 for which \citet{Nattila:2017wtj} has derived: $M = 1.9 \pm 0.3 ~ M_\odot$ and $R = 12.4 \pm 0.4$ km. Additionally, the mass and radius of the millisecond pulsar PSR J0030+0451 have been estimated by the NICER collaboration through pulse profile modelling resulting in: $M = 1.34^{+0.15}_{-0.16} ~ M_\odot$, $R = 12.71^{+1.14}_{-1.19}$ km \citep{riley19c}. Following \cite{Jiang:2019rcw} for both 4U 1702--429 and PSR J0030+0451, we use a bivariate Gaussian distribution to mimic the mass-radius posteriors of these sources:

\begin{equation}
    P(M, R) = \frac{1}{2\pi \sigma_M \sigma_R \sqrt{1-\rho^2}} \exp{\{ -\frac{1}{2(1-\rho^2)} [\frac{(M-\mu_M)^2}{\sigma_M^2}-2\rho \frac{(M-\mu_M)(R-\mu_R)}{\sigma_M \sigma_R} +\frac{(R-\mu_R)^2}{\sigma_R^2}]\}}, 
    \label{eq:mr_post}
\end{equation}

For PSR J0030+0451, we use $\mu_M = 1.34 \rm M_{\odot}$, $\mu_R = 12.71 \rm km$, $\sigma_M = 0.155 \rm M_{\odot}$, $\sigma_R = 1.165 \rm km$, and $\rho = 0.9$. The value of $\rho$ is chosen to mimic the highly correlated behavior of the data \citep{Jiang:2019rcw}. 

Similarly, for 4U 1702--429, we use $\mu_M = 1.9 \rm M_{\odot}$, $\mu_R = 12.4 \rm km$, $\sigma_M = 0.3 \rm M_{\odot}$, $\sigma_R = 0.4 \rm km$, and $\rho = 0.9$, as before to represent the correlation between the measurements. 

Finally, we include the EOS insensitive posterior samples computed by the LVC for the masses and the radii of the two components of GW170817 \citep{TheLIGOScientific:2017qsa,Abbott:2018exr}. \footnote{The data from GW170817 are available at \href{https://dcc.ligo.org/LIGO-P1800115/public}{https://dcc.ligo.org/LIGO-P1800115/public}.}

While this work was in progress, another BNS merger event, GW190425, has been reported by the LVC \citep{Abbott:2020uma}. Since this event is reported to be less constraining of the NS properties, we do not include the data from GW190425 in the present analysis.

\section{Simulation Results}\label{result}
We construct the joint PDFs of the nuclear empirical parameters $\{m^*, K, n_0, S, E_0\}$ following the method described in the previous section assuming the five different priors listed in Table \ref{tab:prior} and including the sources in Section \ref{sources}. Then, we calculate the slope parameter of the symmetry energy ($L$) and the distributions of the RMF coupling constants \textit{a posteriori} from the samples of the PDF. The parameter values corresponding to the most probable points of the joint PDFs are listed in Table \ref{tab:mode_NPP}. Notice that we are including the calculations with hyperons using the baseline and the Wide\_unif priors. Hence, we have seven different calculations for the comparative analyses of the priors and for the effects of the hyperons. 
 A comment about the values of $L$ is mandatory: the most probable values listed in Table \ref{tab:mode_NPP} are typically outside the now accepted range for $L$, see \cite{Xie:2019sqb,Lattimer:2012xj}. The case of the uniform prior is the one with the smallest values of $L$ with its second mode providing $L \sim 50$ MeV. 
Therefore we retain results obtained within the uniform prior to be more relevant from a phenomenological point of view. We notice however that for what concerns astrophysical observables, such as  $R_{1.4}$, they could be related not only to $L$ \cite{Hornick:2018kfi} but also to higher order terms in the symmetry energy expansion, such as $K_{\text{sym}}$ and $J_{\text{sym}}$ \citep{Zhang:2018vbw}, which are however affected by large uncertainties. In particular $-400$MeV $<K_{\text{sym}}<100$MeV and $-200$MeV$<J_{\text{sym}}<800$ MeV, \citep{Zhang:2018vbw}, see \cite{Carson:2018xri} for more recent and tighter constraints.
In the appendix, we show in Figure \ref{neutronmatter} the density dependence of the symmetry energy for the same EOSs together with the values of these two high order derivatives. Interestingly their values are within the uncertainties mentioned above.

\begin{table}[t!]
    \begin{tabular}{|c|c|c|c|c|c|c|}
    \hline
    \hline 
    Models & $m^{\ast}$ & K (MeV) & $n_{0}$ (fm$^{-3}$) & S (MeV) & $E_0$ (MeV) & L (MeV) \\
    \hline 
    Baseline  & $0.759$ & $204.1$ & $0.160$ & $29.3$ & $-16.32$ & $81.2$ \\
    \hline
    Marg\_unif  & $0.710$ & $219.1$ & $0.152$ & $31.3$ & $-16.26$ & $90.9$ \\
    \hline 
    Marg\_Gauss  & $0.713$ & $163.8$ & $0.150$ & $34.5$ & $-16.23$ & $100.6$\\
    \hline 
    Wide\_unif  & $0.760$ & $178.1$ & $0.156$ & $27.0$ & $-16.65$ & $74.7$ \\
                & $0.761$ & $280.2$ & $0.173$ & $19.2$ & $-16.43$ & $49.8$ \\
    \hline 
    Wide\_Gauss  & $0.760$ & $177.6$ & $0.151$ & $37.6$ & $-16.16$ & $106.9$  \\
    \hline 
    Wide\_unif with $\Lambda$  & $0.682$ & $319.8$ & $0.170$ & $17.6$ & $-16.72$ & $50.6$  \\
    \hline 
    Baseline with $\Lambda$ & $0.704$ & $279.2$ & $0.142$ & $28.7$ & $-16.11$ & $83.2$  \\
    \hline
    \end{tabular}
    \caption{Most probable empirical parameters from the joint posterior along with the calculated $L$.  Additionally, for the Wide\_unif the parameters associated to the second mode of the PDF are also listed.}
    \label{tab:mode_NPP}
\end{table}

We use the python \textsf{corner.py} package to visualize one- and two-dimensional projection plots of the samples \citep{corner}. In the two-dimensional plots we show the contours at $1\sigma$ (39.3\%), 68\% and 90\% confidence interval (CI). Next, we draw the mass-radius sequences corresponding to the samples within $68\%$ CI of the joint posterior and include the most likely sequences for all the priors. The maximum mass and the $R_{1.4}$ for the most likely sequences along with the minimum and maximum values of $R_{1.4}$ and $M_{max}$ calculated from the samples within the 68\% CI of the joint posterior are listed in Table \ref{tab:m-r14}. The median values of the marginalized distribution of $R_{1.4}$ are listed in Table \ref{tab:median_r14}.  It should be noted that the values of the most probable sequence are associated with the most probable set empirical parameters of the joint posterior. This does not necessarily correspond to the distribution of the radii calculated from the those EOSs. For example, several sets of parameters which give equally big radii can shift the histogram of the radii towards larger values. Therefore, even though those combinations are less probable individually, they shift the peak of the radius distribution away from the value corresponding to the most probable set. 

We calculate the evidence and BIC for each model and compute the Bayes factors (BF) as the ratio between evidences. This allows for a quantitative comparison on the best plausible scenario among the ones considered in this study. The values of the BFs and the $\Delta$BICs are listed in Tables \ref{tab:bayes_factor} and \ref{tab:bic} respectively. For the sake of clearness the tables contain all the seven cases, but in the main text only the plots and results for baseline, Wide\_unif, Wide\_Gauss priors and baseline prior with $\Lambda$ are presented. We refer to the appendix for the details of the calculations and the results obtained for Marg\_unif, Marg\_Gauss priors and Wide\_unif prior with $\Lambda$.

\begin{table}[b!]
    \centering
    \begin{tabular}{|c|c|c|c|c|c|c|c|}
    \hline
    \hline 
    Models & Baseline & Marg\_unif & Marg\_Gauss & Wide\_unif & Wide\_Gauss & \pbox{1.5cm}{Wide\_unif \\ with $\Lambda$} & \pbox{1.5cm}{Baseline \\ with $\Lambda$} \\
    \hline 
    $R_{1.4}$ (Km) & $12.58$ & $13.47$ & $12.55$ & $11.70$ \vline $\,\,\,\, 12.15$ & $12.32$ & $12.80$ & $14.18$\\
    \cline{1-8}
    $\Delta R_{1.4}$ (Km) & $12.48 - 13.30$ & $13.43 - 13.84$ & $11.82 - 13.24$ & $11.16 - 12.78$ & $11.32 - 13.18$ & $12.71 - 13.25$ & $14.13 - 14.24$\\
    \cline{1-8}
    $M_{max}$ (M$_{\odot}$) & $2.01$ & $2.29$ & $2.27$ & $2.00$ \vline $\,\,\,\, 2.00$ & $2.03$ & $2.01$ & $2.00$\\
    \cline{1-8}
    $\Delta M_{max}$ (M$_{\odot}$) & $2.00 - 2.24$ & $2.28 - 2.42$ & $2.11 - 2.46$ & $2.00 - 2.39$ & $2.00 - 2.40$ & $2.00 - 2.10$ & $2.00 - 2.03$\\
    \cline{1-8}
    \hline
    \end{tabular}
    \caption{$R_{1.4}$ and maximum mass of the most probable configuration of the joint posterior.  Ranges of $R_{1.4}$ and maximum mass corresponding to the samples within $68\%$ CI of the empirical parameters.}
    \label{tab:m-r14}
\end{table}

\begin{table}[b!]
    \centering
    \begin{tabular}{|c|c|c|c|c|c|c|c|}
    \hline
    \hline 
    Models & Baseline & Marg\_unif & Marg\_Gauss & Wide\_unif & Wide\_Gauss & Wide\_unif with $\Lambda$ & Baseline with $\Lambda$  \\
    \hline 
    $R_{1.4}$ (Km) & $12.95_{-0.20}^{+0.25}$ & $13.68_{-0.12}^{+0.09}$ & $12.56_{-0.29}^{+0.37}$ & $12.42_{-0.33}^{+0.31}$ & $12.37_{-0.38}^{+0.37}$ & $13.00_{-0.16}^{+0.22}$ & $14.19_{-0.03}^{+0.03}$ \\
    \hline
    \end{tabular}
    \caption{Median of the distribution for $R_{1.4}$.}
    \label{tab:median_r14}
\end{table}

\begin{figure}[h!]
    \centering
    \includegraphics[height=0.8\textheight]{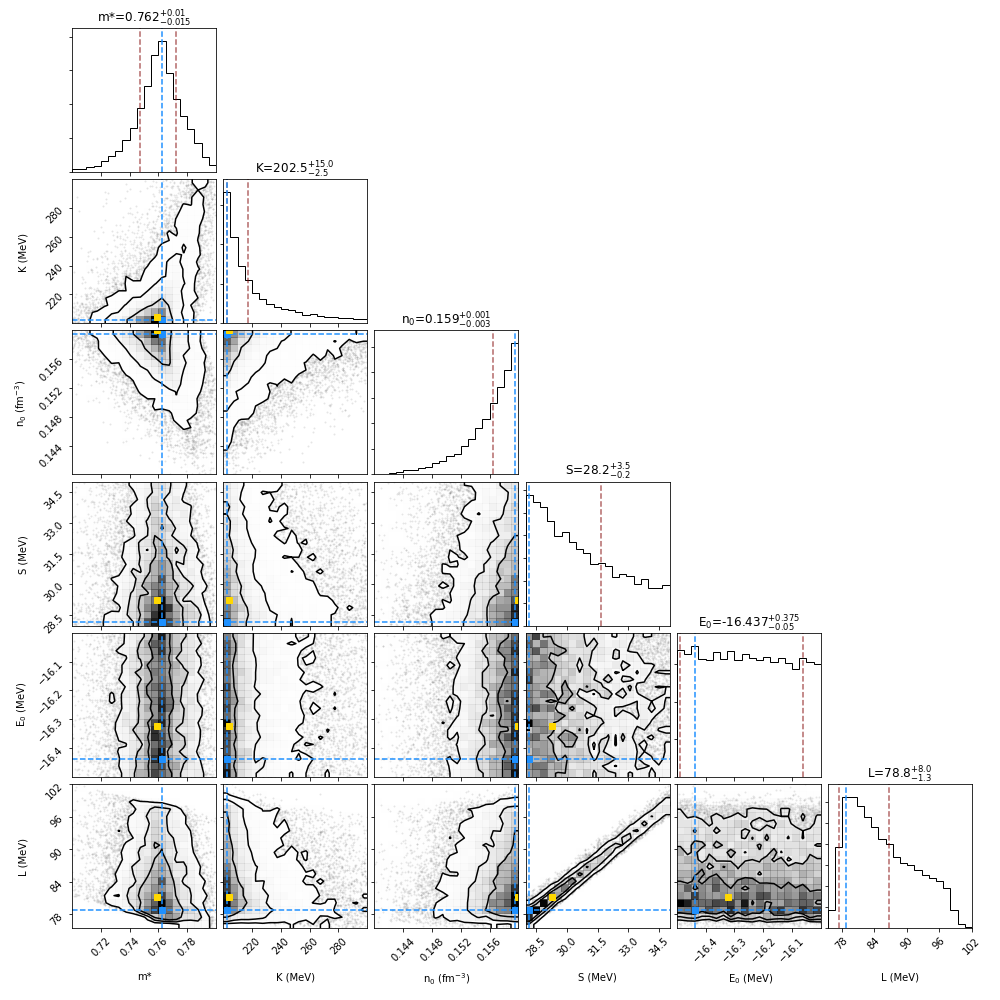}   
    \caption{Posterior distributions of the empirical parameters for the baseline prior. In the marginalized one-dimensional plots the blue lines correspond to the mode and the maroon lines the $1\sigma$ CI. The contours in the two-dimensional PDFs are at $1\sigma$ (39.3\%), 68\% and 90\% CI respectively. The yellow points represent the most probable values for the joint posterior. }
    \label{NPP_unif_mode}
\end{figure}

\begin{figure}[h!]
    \centering
        \begin{tabular}{cc}
         \includegraphics[width=0.5\textwidth]{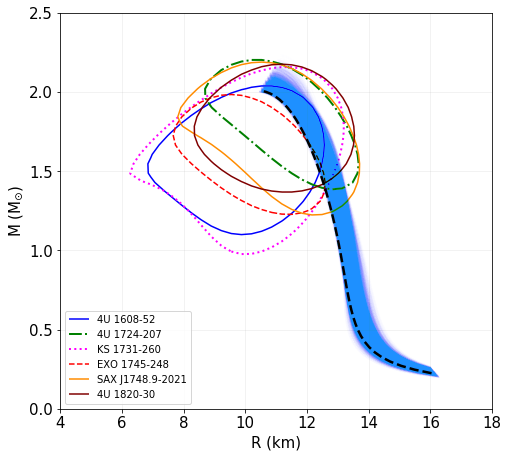}& \includegraphics[width=0.5\textwidth]{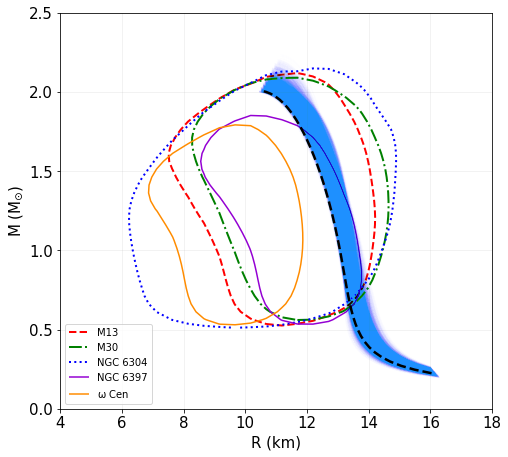} \\
        \includegraphics[width=0.5\textwidth]{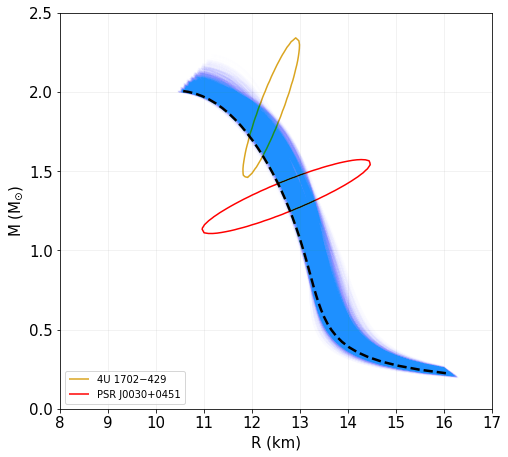} & 
        \includegraphics[width=0.5\textwidth]{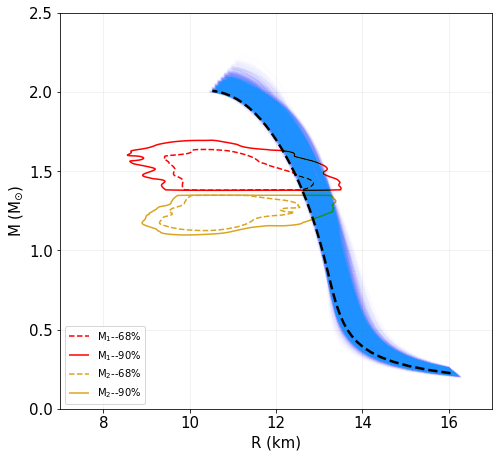}\\
    \end{tabular}
        \caption{Inferred mass-radius curves corresponding to the EOS parameters up to 68\% CI assuming the baseline prior along with the sources. The bottom right panel corresponds to the event GW170817. The black dashed lines in all the panels represent the most probable EOS parameter set.}
    \label{fig:unif_mr}
\end{figure}

In Figure \ref{NPP_unif_mode}, we present the marginalized PDFs for the empirical parameters with the baseline prior (the corresponding RMF coupling constants are shown in Figure \ref{RMFP_unif_mode} in the appendix). The most probable parameters of the joint PDF are shown as the yellow points in the marginalized plots. We see a prominent peak for $m^*$, a flat distribution in $E_0$, while $K$, $n_0$, and $S$ show the trends of having the most probable configurations at the edge of the prior boundaries, respectively. The peak of $m^*$ can be understood by considering its strong correlation with the maximum mass \citep{Weissenborn:2011kb}, the smaller the value of $m^*$, the larger the value of the maximum mass. Since at the same time, the sources point towards not-too-large radii, $m^*$ prefers values which are not too small. For what concerns $K$, $n_0$, and $S$, they affect the stiffness of the EOS at not too high densities and therefore their values are mostly constrained by the radii of the sources. This can be understood more clearly from the inferred mass-radius curves plotted in Figure \ref{fig:unif_mr}. The most probable sequence lies on the edge of the 68\% CI of the inferred EOSs. It is also compatible at the 68\% CI for most of the sources. Only  $\omega$Cen shows a mild tension with the most probable sequence.

\begin{figure}[h!]
    \centering
    \includegraphics[height=0.8\textheight]{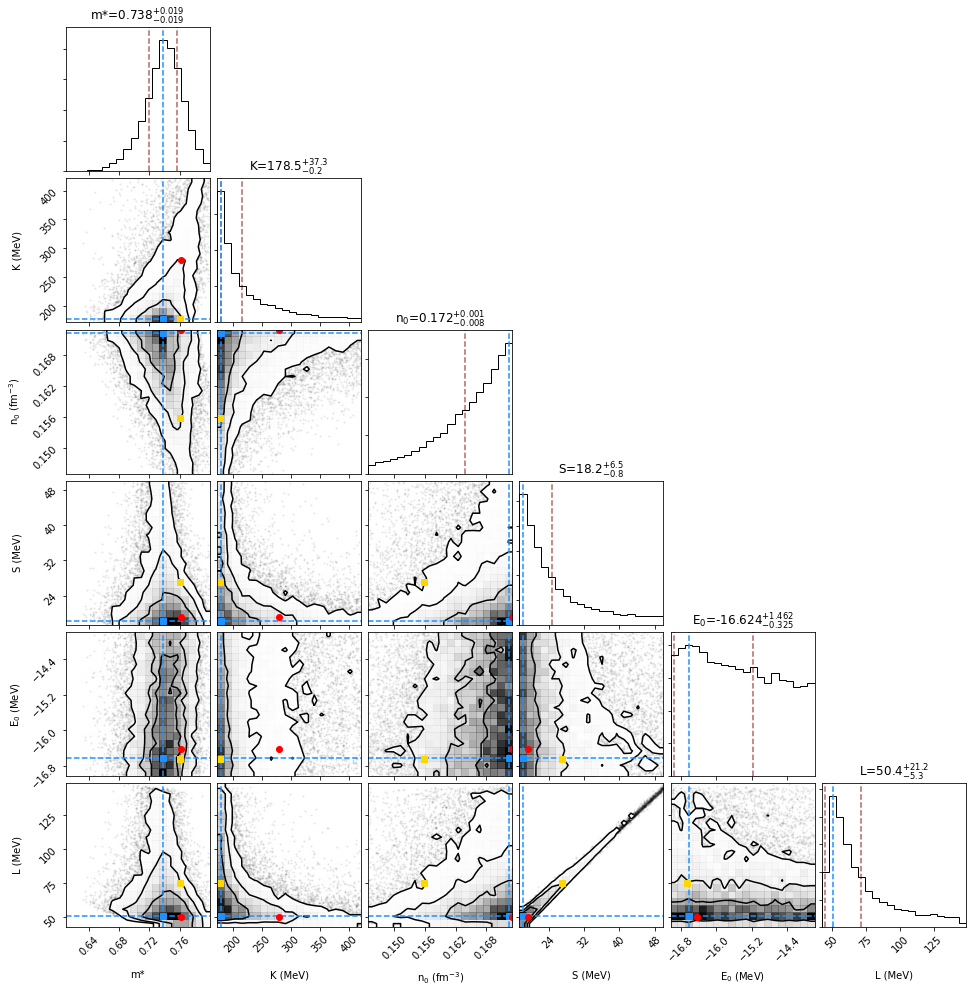}   
    \caption{Same as Figure \ref{NPP_unif_mode} for the Wide\_unif prior. The red points indicate the second mode of the joint distribution.}
    \label{NPP_unif_wide_mode}
\end{figure}

\begin{figure}[h!]
    \centering
        \begin{tabular}{cc}
         \includegraphics[width=0.5\textwidth]{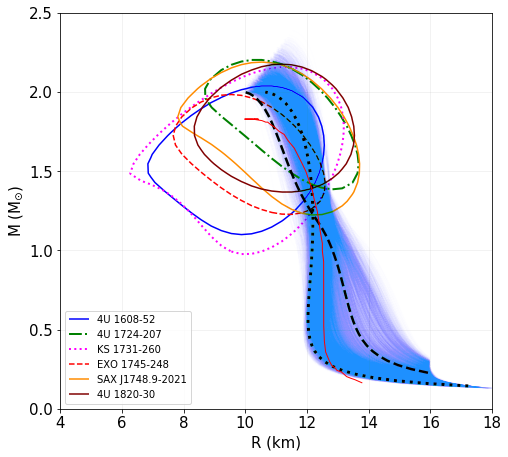}& \includegraphics[width=0.5\textwidth]{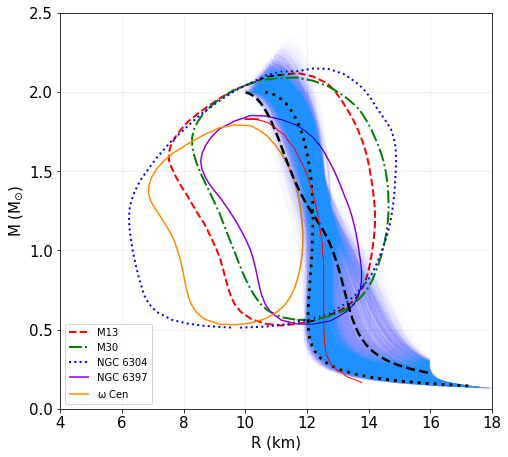} \\
        \includegraphics[width=0.5\textwidth]{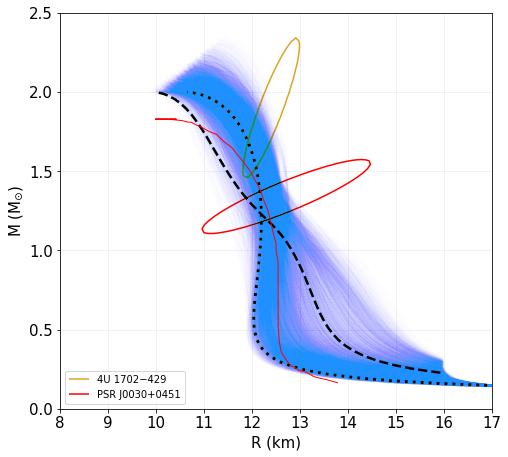} & 
        \includegraphics[width=0.5\textwidth]{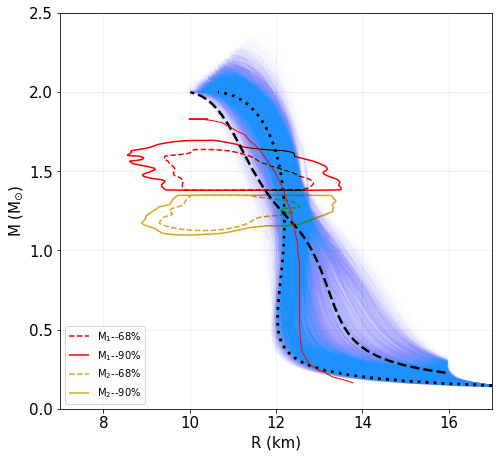}\\
    \end{tabular}
        \caption{Same as Figure \ref{fig:unif_mr} for the Wide\_unif prior. Here the black dotted lines in all the panels represent the curve related to the second mode of the joint distribution for the empirical parameters. The red curve shows LS180 for comparison.}
    \label{fig:unif_wide_mr}
\end{figure}

In Figure \ref{NPP_unif_wide_mode} , we present the result for the empirical parameters with the Wide\_unif prior. The $m^*$ is peaked at a slightly different value than the baseline prior. The other parameters $K$, $n_0$, and $S$ also follow a similar trend with the most probable values at the edges of the parameter space. In the joint posterior, we find a hint of bimodality which is not evident in the marginalized PDFs. In Figure \ref{NPP_unif_wide_mode}, we indicate the values corresponding to the two modes with yellow (absolute maximum) and blue (second relative maximum) points. From Table \ref{tab:mode_NPP}, we see that the value of $S$ for the second maximum is most likely ruled out by presently available nuclear physics analysis. On the other hand, the most probable parameter set is striking similar to that of the LS180 supernova EOS, whose values for the empirical parameters are $K=180$ MeV, $n_0=0.155 fm^{-3}$, $S=28.6$ MeV and $L=73.8$ MeV \citep{LATTIMER1991331}. It is remarkable that a RMF model can reproduce the same results obtained within a medium-dependent liquid-drop model.

Concerning the the M-R sequences in Figure \ref{fig:unif_wide_mr}, they also feature a bimodal behavior resulting from the joint PDF. One can see two different regions in the M-R plots where the sequences are clubbed together. The dashed and the dotted lines correspond to the two modes respectively and show a good agreement with the astrophysical data. However the dotted line is most likely ruled out as explained before \footnote{ Notice that this M-R curve is quite similar to the result of the APR EOS \cite{APR} which however is perfectly compatible with the present knowledge of nuclear symmetry energy. Again, introducing in our Lagrangian other terms aiming at a better description of symmetry energy would possibly lead to solutions like the dashed line which are consistent with nuclear physics data and astrophysical data.}. Interestingly, the most probable sequence shows a bump close to $\sim 1 M_\odot$ and contrary to LS180, it reaches the $2 M_\odot$ limit and its $R_{1.4}=11.7$ km thus $0.5$ km smaller than the LS180 value. While the saturation properties of these two EOSs are very similar, they show different high density behavior for $\beta-$stable matter. This can be interpreted as the outcome of the nonlinearity of the RMF parametrization which will be discussed in the next section.

\begin{figure}[h!]
    \centering
    \includegraphics[height=0.8\textheight]{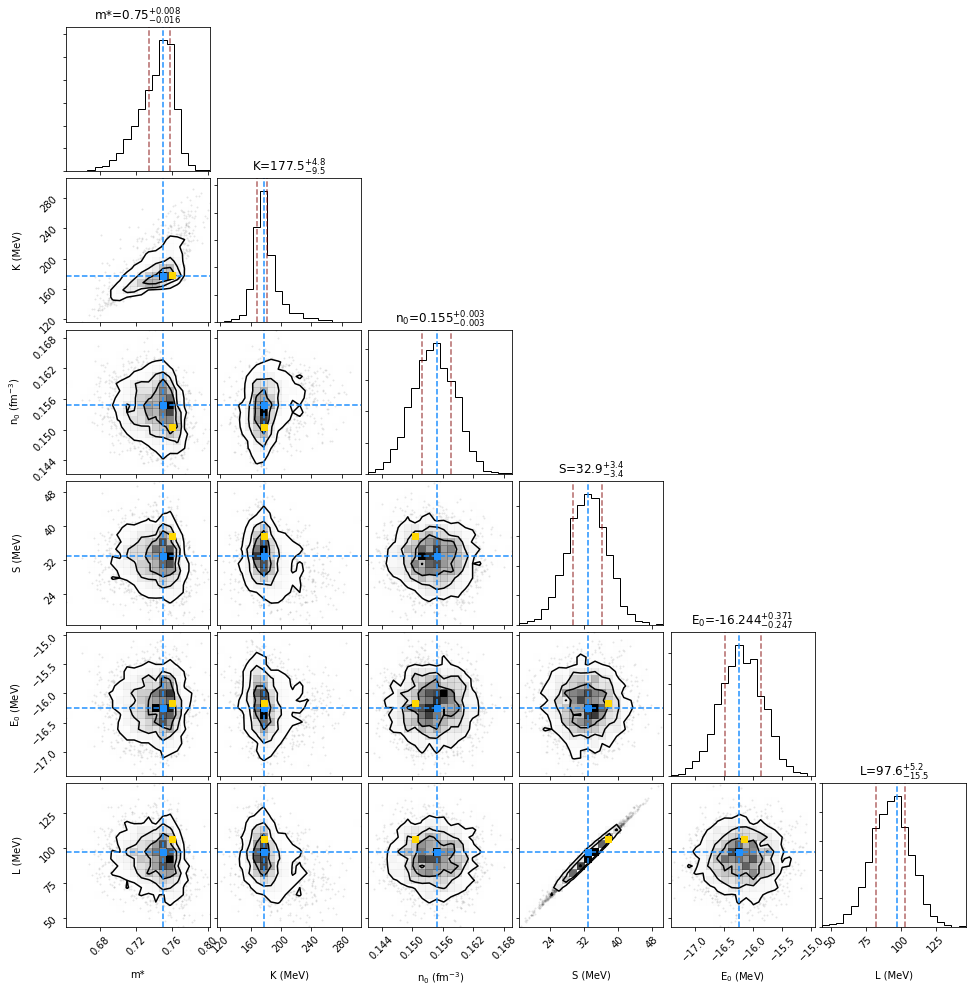}   
    \caption{Same as Figure \ref{NPP_unif_mode} for the Wide\_Gauss prior}
    \label{NPP_gau_wide_mode}
\end{figure}

\begin{figure}[h!]
    \centering
        \begin{tabular}{cc}
         \includegraphics[width=0.5\textwidth]{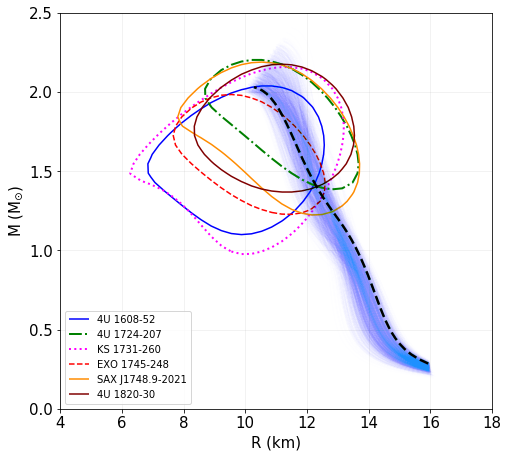}& \includegraphics[width=0.5\textwidth]{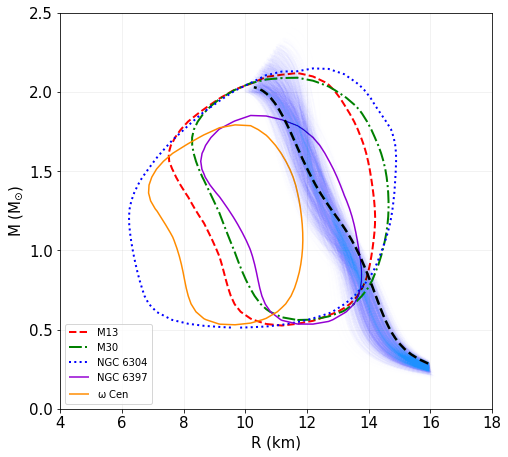} \\
        \includegraphics[width=0.5\textwidth]{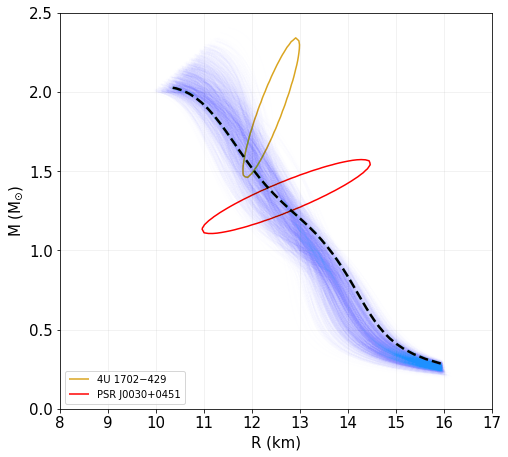} & 
        \includegraphics[width=0.5\textwidth]{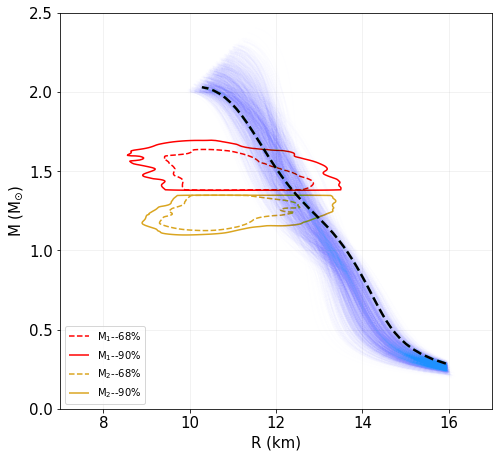}\\
    \end{tabular}
        \caption{Same as Figure \ref{fig:unif_mr} for the Wide\_Gauss prior.}
    \label{fig:gau_wide_mr}
\end{figure}

In Figure \ref{NPP_gau_wide_mode}, the results for the Wide\_Gauss prior are shown. Apart from $K$, the posteriors for the other parameters do not deviate much from the prior distribution. We see a significant shift for the peak of $K$, which moves about $2\sigma$ below the peak of the prior. 
It implies that the EOS prefers to be rather soft at saturation but stiffer at the higher density due to the high symmetry energy and not too high effective mass. A small value of $K$ would be consistent with the analysis of the KaoS collaboration on heavy ions collisions \citep{Sagert:2011kf}. The effect of using a Gaussian prior is to remove the bimodality found in the previous case but the bump associated with the mass-radius curve of the most probable EOS is still present as can be seen in Figure  \ref{fig:gau_wide_mr}. The value of $R_{1.4}$ is again smaller than the value obtained with the baseline prior. The correlation between $K$ and the radii has also been noted in \cite{Ferreira:2019bny}.

\begin{figure}
    \centering
    \includegraphics[height=0.8\textheight]{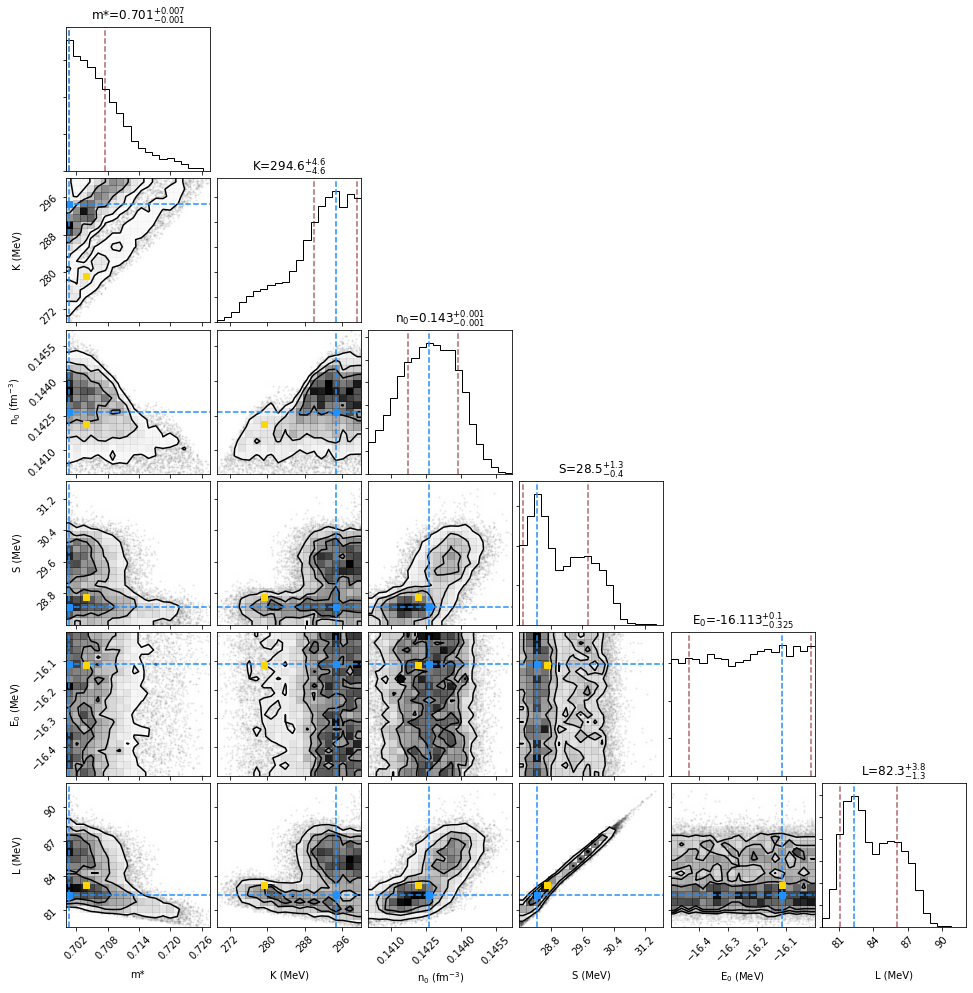}   
    \caption{Same as Figure \ref{NPP_unif_mode} for the baseline prior with $\Lambda$ in the system.}
    \label{NPP_lam_glen_mode}
\end{figure}

\begin{figure}
    \centering
        \begin{tabular}{cc}
         \includegraphics[width=0.5\textwidth]{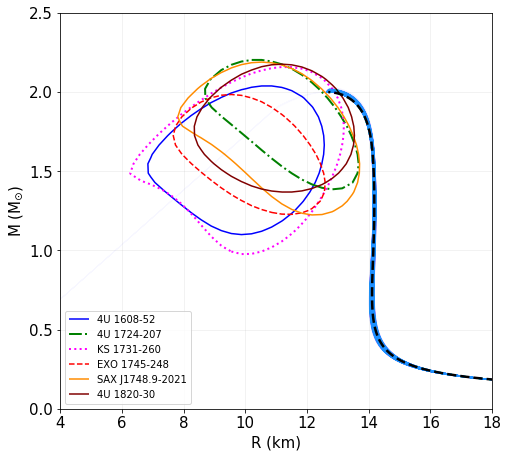}& \includegraphics[width=0.5\textwidth]{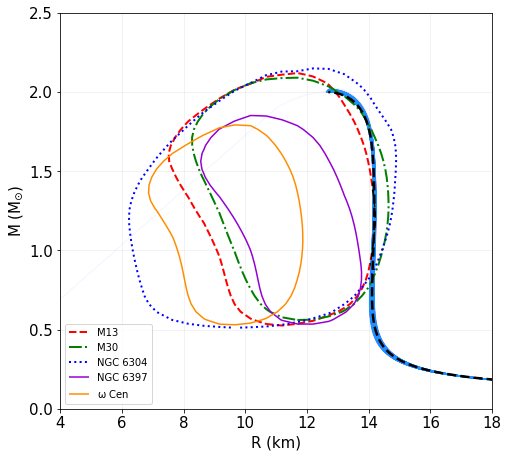} \\
        \includegraphics[width=0.5\textwidth]{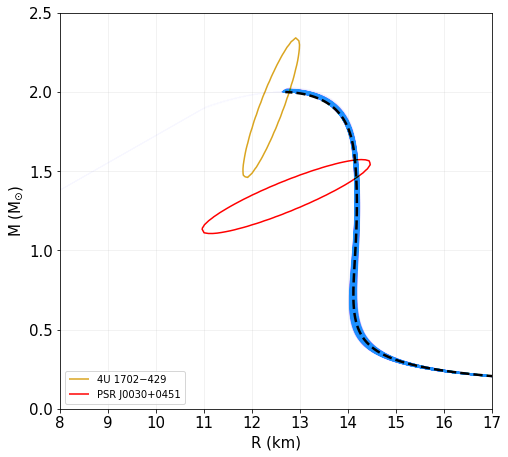} & 
        \includegraphics[width=0.5\textwidth]{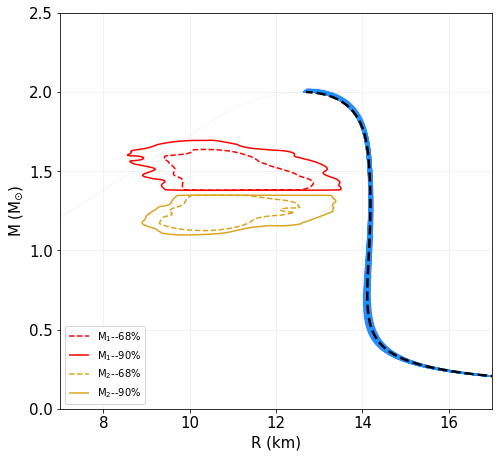}\\
    \end{tabular}
        \caption{Same as Figure \ref{fig:unif_mr} for the baseline prior range with $\Lambda$ hyperon in the system.}
    \label{fig:lam_glen_mr}
\end{figure}

Next, we investigate the effect of the formation of $\Lambda$ hyperons in the system. As expected, the appearance of a new degree of freedom softens the EOS. Therefore most parts of the parameter space are not consistent with the $2M_\odot$ limit and are ruled out. This restricts the parameter space severely. The posteriors for the empirical parameters with baseline prior and the corresponding  M-R sequences are shown in Figures \ref{NPP_lam_glen_mode} and \ref{fig:lam_glen_mr} respectively. The qualitative differences with respect to the previous cases are evident in the marginalized PDF of $m^*$ and $K$. In particular, $m^*$ does not exhibit a peak and its distribution tends towards the lower edge of the range. Whereas $K$ prefers the higher edge contrary to the previous cases. These behaviors arise to compensate the softening associated with the hyperons in order to fulfil the $2 M_\odot$ criterion. We get M-R sequences which are outside the 68\% regions of the posteriors of most of the sources. The value of $R_{1.4}$ for the most probable EOS is more than 14km, significantly larger than the other cases. In the case of the Wide\_unif prior, we find some sequences within the observable limits but the symmetry energy turns out to be  very small, well outside the acceptable region (See Figures \ref{NPP_lam_mode} and \ref{fig:lam_mr} in the appendix).

Finally, we compare our models using BIC and evidence calculations. From Table \ref{tab:bayes_factor}, we find the highest evidence for the Wide\_Gauss prior and the lowest for the Marg\_unif prior for the nucleonic models. Between the hyperonic models, the Wide\_unif is preferred over the baseline prior. In general, nucleonic models are preferred over the hyperonic models. One  can draw the same conclusion from the calculations of $\Delta$BIC in Table \ref{tab:bic}, the only exception being the Marg\_Gauss case. The calculation of BIC includes only the likelihood for the most probable configuration and the number of prior parameters. Instead, the BF takes into account the full prior distribution. Therefore, in the case of Gaussian priors the probabilities away from central value are suppressed. So, two different models with parameters producing similar maximum likelihood values can have similar BIC but different BF if the nature of the prior distribution is different. This case is realized in the Marg\_Gauss prior, as the peak of the posterior for $K$ deviates by more than $3\sigma$ w.r.t the peak of the prior. 

Now, let us focus on the interpretation of the BFs between the two categories of priors: the informed and the agnostic. For the informed prior, we take the baseline model and for the agnostic the Wide\_unif model, for example. If the maximum likelihood points for both of the models happen to be inside the overlap region of the priors, one would expect the evidence for the Wide\_unif to be smaller. This is due to its larger prior volume  penalizing it over the baseline prior. The fact that the Wide\_unif has a larger evidence instead suggests a tension between the nuclear physics informed priors and the astrophysical data, consistent with the findings of \cite{Guven:2020dok}.




\begin{table}[h!]
    \centering
    \begin{tabular}{| c @{\hskip 1pt} |  c | c | c | c | c | c | c |}

    \hline
    \hline 
      \backslashbox{Model 1}{Model 0} & Baseline & Marg\_unif & Marg\_Gauss & Wide\_unif & Wide\_Gauss &  \pbox{1.5cm}{Wide\_unif \\ with $\Lambda$} & \pbox{1.5cm}{Baseline \\ with $\Lambda$} \\

    \hline 
    Baseline  &  & $4 \cdot 10^{-5}$ & $0.04$ & $6.5$ & $30.2$ & $0.03$ & $\mathbf{1.4 \cdot 10^{-7}}$\\
            & & n. decisive & n. strong & p. moderate & p. strong & n. very strong & \textbf{n. decisive} \\
    \hline
    Marg\_unif  & $2.4 \cdot 10^{4}$ &  & $951.3$ & $1.6 \cdot 10^{5}$ & $7.2 \cdot 10^{5}$ & $710.4$ & $0.003$ \\
            & p. decisive & & p. decisive & p. decisive & p. decisive & p. decisive & n. decisive\\
    \hline 
    Marg\_Gauss  & $25.1$ & $0.001$ &  & $163.7$ & $756.8$ & $0.75$ & $3.6 \cdot 10^{-6}$ \\
            & p. strong & n. decisive &  & p. decisive & p. decisive & n. weak & n. decisive \\
    \hline 
    Wide\_unif  & $\mathbf{0.15}$ & $6.4 \cdot 10^{-6}$ & $0.006$ &  & $4.6$ & $\mathbf{0.005}$ & $2.2 \cdot 10^{-8}$ \\
            & \textbf{n. moderate} & n. decisive & n.decisive &  & p. moderate & \textbf{n. decisive} & n. decisive\\
    \hline 
    Wide\_Gauss  & $0.03$ & $1.4 \cdot 10^{-6}$ & $0.001$ & $\mathbf{0.21}$ &  & $0.001$ & $4.7 \cdot 10^{-9}$ \\
             & n. strong & n. decisive & n. decisive & \textbf{n. moderate} &  & n. decisive & n. decisive \\
    \hline 
    Wide\_unif   & $33.6$ & $0.001$ & $1.34$ & $219.2$ & $1013.4$ & & $4.8 \cdot 10^{-6}$ \\
    with $\Lambda$ & p. very strong & n. decisive & p.weak & p. decisive & p.decisive & & n. decisive \\
    \hline    
    Baseline  & $7. \cdot 10^6$  & $293.5$ & $2.8 \cdot 10^{5}$ & $4.6 \cdot 10^{7}$ & $2.1 \cdot 10^{8}$ & $2.1 \cdot 10^{5}$ &  \\
    with $\Lambda$ & p. decisive & p. decisive & p. decisive & p. decisive & p.decisive & p. decisive & \\
    \hline    
    \end{tabular}
    \caption{Bayes factors (BF$_{01}$). We indicate the strength of the preference between the two compared models. Here p. (n.) suggest a positive (negative) preference for model 0 over model 1.   }
    \label{tab:bayes_factor}
\end{table}

\begin{table}[h!]
    \centering
    \begin{tabular}{| c  @{\hskip 1pt} | c | c | c | c | c | c | c |}
    \hline
    \hline 
      \backslashbox{Model 1}{Model 0} & Baseline & Marg\_unif & Marg\_Gauss & Wide\_unif & Wide\_Gauss &  \pbox{1.5cm}{Wide\_unif \\ with $\Lambda$} & \pbox{1.5cm}{Baseline \\ with $\Lambda$} \\    
    \hline 
    Baseline  &  & $17.72$ & $-1.07$ & $-2.11$ & $-2.33$ & $2.49$ & $\mathbf{24.00}$\\
            & & n. decisive & p. weak & p. moderate & p. moderate & n. moderate & \textbf{n. decisive}\\
    \hline
    Marg\_unif  & $-17.72$ &  & $-18.79$ & $-19.83$ & $-20.05$ & $-15.24$ & $6.26$\\
            & p. decisive & & p. decisive & p. decisive & p. decisive & p. decisive & n. strong \\
    \hline 
    Marg\_Gauss  & $1.07$ & $18.79$ &  & $-1.04$ & $-1.26$ & $3.55$ & $25.05$ \\
            & n. weak & n. decisive &  & p. weak & p. weak & n. moderate & n. decisive\\
    \hline 
    Wide\_unif  & $\mathbf{2.11}$ & $19.83$ & $1.04$ &  & $-0.22$ & $\mathbf{4.60}$ & $26.10$\\
            & \textbf{n. moderate} & n. decisive & n. weak &  & p. weak & \textbf{n. moderate} & n. decisive \\
    \hline 
    Wide\_Gauss  & $2.33$ & $20.05$ & $1.26$ & $\mathbf{0.22}$ &  & $4.81$ & $26.31$ \\
             & n. moderate & n. decisive & n. weak & \textbf{n. weak} &  & n. moderate & n. decisive \\
    \hline 
    Wide\_unif   & $-2.49$ & $15.24$ & $-3.55$ & $-4.60$ & $-4.81$ & & $21.50$ \\
    with $\Lambda$ & p. moderate & n. decisive & p. moderate & p. moderate & p. moderate & & n. decisive \\
    \hline   
    Baseline  & $-24.00$ & $-6.26$ & $-25.05$ & $-26.10$ & $-26.31$ & $-21.50$ & \\
    with $\Lambda$ & p. decisive & p. strong & p. decisive & p. decisive & p. decisive & p. decisive &  \\
    \hline   
    \end{tabular}
    \caption{$\Delta$BIC$_{01}$. We indicate the strength of the preference between the two compared models. Here p. (n.) suggest a positive (negative) preference for model 0 over model 1.}
    \label{tab:bic}
\end{table}
 
\section{Discussion and conclusions}\label{discussion}

We have shown before that when considering the uniform informed priors the M-R relations are qualitatively very similar to the old parameterization of \cite{Glendenning:1991es}. This result leads to the following conclusion: within the present uncertainties on the nuclear physics empirical parameters and by using uniform priors, it is basically impossible to obtain $R_{1.4}$ smaller than $\sim 12.5$km, see Tables \ref{tab:m-r14} and \ref{tab:median_r14} and Figure \ref{fig:unif_mr}. Radii close to $\sim 11$km can be reached only for configurations close to the maximum mass.  This conclusion is clearly based on our modelling of the EOS and specifically on our choice of relativistic mean field model. However we would like to remark that this outcome 
 is fully in agreement with the results of the metamodelling analysis of  \citet{Margueron:2017lup} and the independent analysis by \citet{Most:2018hfd}.  In both those papers the equation of state is modelled in a different way: a Taylor expansion around saturation in the first case, a chiral effective field theory up to densities close to saturation and a piecewise polytropic parametrization for larger densities.
On the other hand, the fact that the most probable M-R curve sits in both cases on the left border of the $68\%$ CI is somehow suggesting that there is some tension between the values of the nuclear physics empirical parameters and the astrophysical measurements, as stressed before.
That is the main reason for investigating different kinds of priors.

\begin{figure}[!ht]
	\begin{centering}
    	\epsfig{file=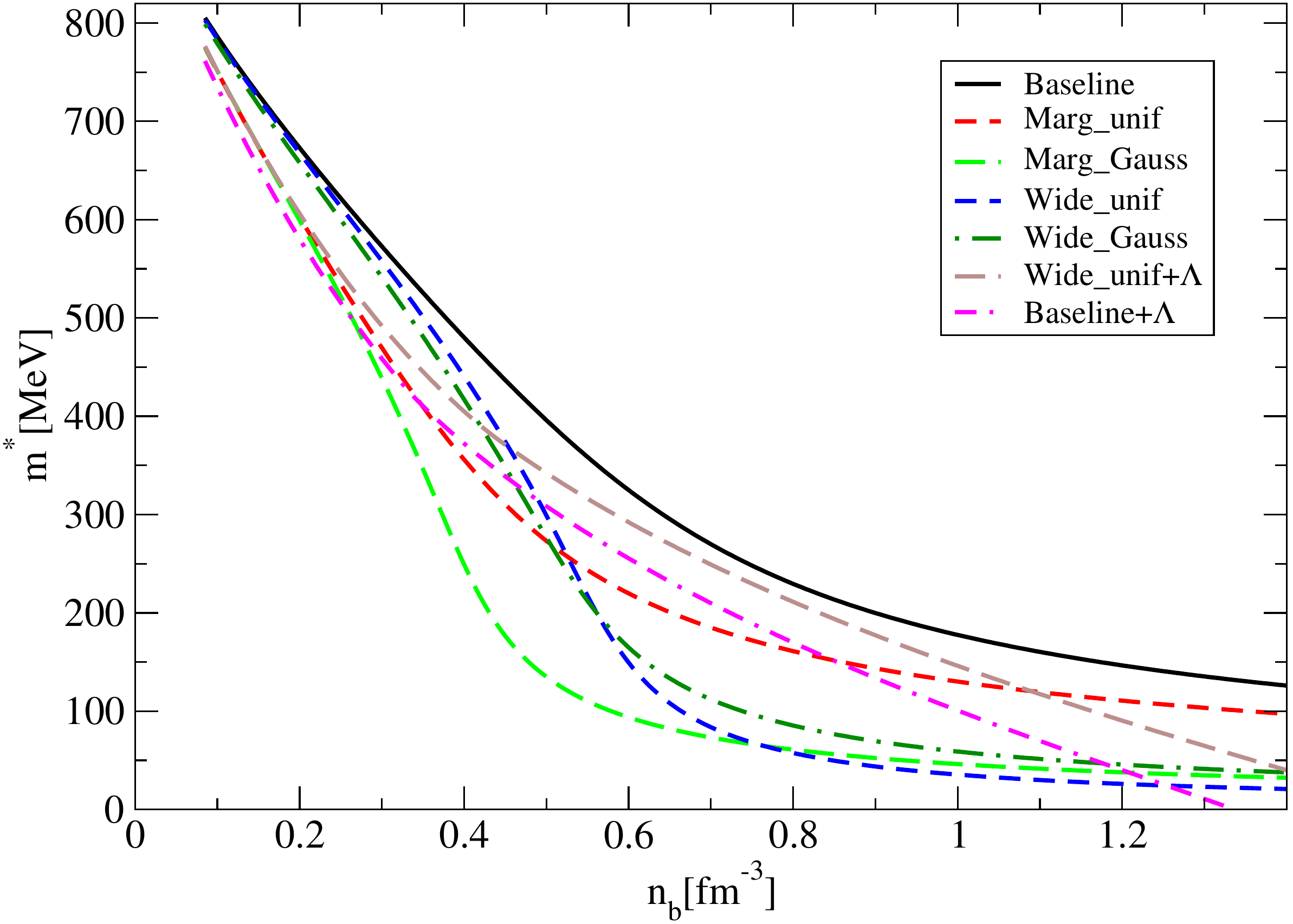,height=7cm,width=11cm}

		\caption{Effective mass as a function of density for the five priors here adopted. The curves correspond to the most probable parameters for each prior. Also the two cases with the inclusion of $\Lambda$ are shown.\label{masseff}}
	\end{centering}
\end{figure}

\begin{figure}[!ht]
	\begin{centering}
		\epsfig{file=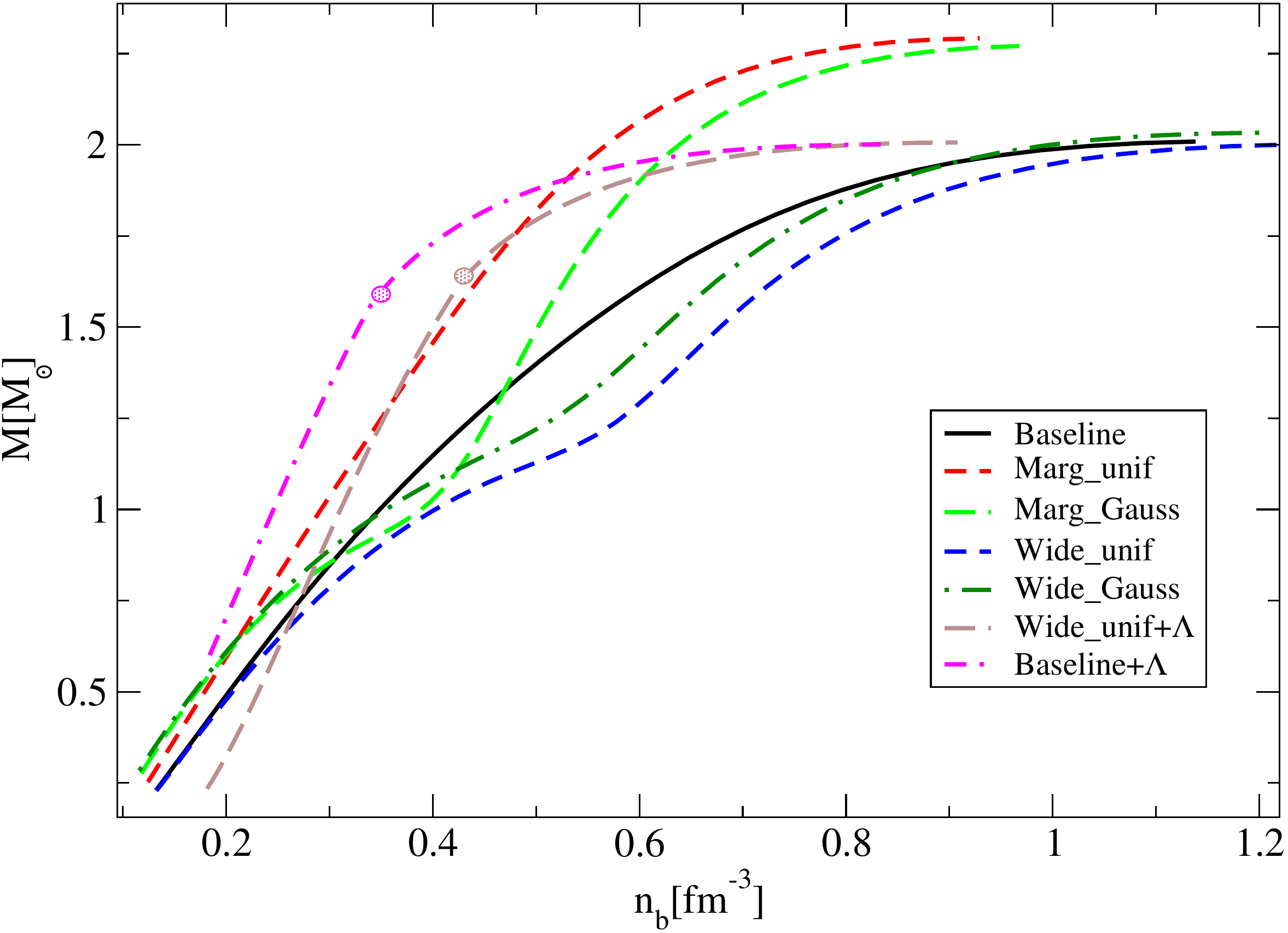,height=7cm,width=11cm}
		\caption{Gravitational mass as a function of the central density as obtained from the same EOSs of Fig.\ref{masseff}. The circles correspond to the onset of $\Lambda$'s in the EOSs. \label{massrhoc}}
	\end{centering}
\end{figure}

Quite remarkably, when allowing for a wider exploration of the parameters (Wide\_unif, Wide\_Gauss and Marg\_Gauss, see appendix), we see a qualitative difference in the M-R relation with a bump appearing at about $1M_{\odot}$. That behaviour is very similar to the one obtained when a smooth phase transition occurs (for instance, a Gibbs mixed phase with quark matter). To understand the origin of this behaviour, one has to check the density dependence of the effective mass (displayed in Figure \ref{masseff}) which, due to the non-linearity of the sigma potential, could feature a phase transition. Indeed, $m^*$ features a fast drop for the last three priors at densities of $2-3$ times saturation density. Moreover also a change in concavity with respect to the first two priors can be noticed. This bump is clearly visible also in the plot showing the gravitational mass as a function of the central baryon density, see Fig.\ref{massrhoc}.
Such a behaviour is reminiscent of a partial restoration of chiral symmetry and we interpret it as a clear indication that astrophysical measurements are in fact suggesting the appearance of new phases of strongly interacting matter in neutron stars. Notice that the prior with the largest evidence is the one adopting wide Gaussian distributions for the nuclear empirical parameters: in this case $R_{1.4}$ could drop below $\sim 12$ km and radii as small as $\sim 10$ km are reached at the maximum mass configuration, see Figure \ref{fig:gau_wide_mr}.

Similar conclusion have been drawn in other works employing different methods: in \citet{Essick:2019ldf,Guven:2020dok,Fujimoto:2019hxv}, hints for a phase transition have been suggested by using the non parametric inference, the metamodelling tools and the deep neural network methods. Previously, also a Bayesian analysis with a tuned parameterization of strong phase transition, has led to the very same conclusion \citep{Steiner:2017vmg}.

Let us discuss now the cases in which also hyperons have been included in the EOS.
In Fig. \ref{massrhoc}, we display the relation between gravitational mass and central density for the different models. In all the cases in which hyperons are not included, the central density for the $2M_{\odot}$ configuration is larger than about $4n_0$. Therefore, to justify the approximation of neglecting hyperons in those EOSs, one has to resort to some stiffening mechanism that shifts the threshold of appearance of hyperons to larger density, e.g. a strong repulsion due to three body interactions \citep{Lonardoni:2014bwa} or hidden strangeness vector field as the $\phi$ meson, see \citet{Weissenborn:2011ut}. On the other hand, when allowing for the formation 
of hyperons (only $\Lambda$'s and with fixed couplings in this work) $\Lambda$'s do appear in compact stars with masses larger than about $1.6M_{\odot}$ and up to $2M_{\odot}$. For the Wide\_unif case, the possibility of forming hyperons and, at the same time, of fulfilling the $2M_{\odot}$ limit is due to an unreasonably small value (both the most probable and the median) of the symmetry energy (see Tables \ref{tab:mode_NPP} and \ref{tab:median_NPP}) and a very large value of $K$. We consider this EOS to be very unlikely. By introducing $\Lambda$'s with the baseline prior, the need of reaching $2M_{\odot}$ causes an early stiffening of the EOS such that $R_{1.4} \gtrsim 14$ km. This result is in agreement with previous findings presented in 
\citet{Fortin:2014mya} where it has been suggested that presence of hyperons systematically shifts the radii of compact stars to larger value with respect to nucleonic stars. As one can notice in Fig. \ref{fig:lam_glen_mr} the predicted mass-radius curve are outside even the $90\%$ CI provided by GW170817. 
It is clear that our minimal modelling for including hyperons strengthens
the tension between the values of the nuclear physics empirical parameters and the astrophysical measurements. This statement is supported by the results on the BFs.

We plan to further investigate this issue in the future by adding for instance a repulsive channel for hyperons. Indeed, there are many examples in the literature in which hyperons do not forbid to obtain values of $R_{1.4}$ close to $12.5-13$km while fulfilling the $2M_{\odot}$ limit, see e.g. \citet{Logoteta:2019utx} for a recent microscopic calculation and \citet{Weissenborn:2011ut,Maslov:2015msa,Negreiros:2018cho} for relativistic mean field examples. Also delta resonances should be in principle included since, as shown in many papers, their onset is actually comparable to the one of hyperons \cite{Drago:2014oja,Cai:2015hya,Li:2018qaw}.
The indications of a  possible phase transition found in this paper,
should also be investigated thoroughly by considering a transition to quark matter via a Gibbs construction, see \citet{Weissenborn:2011qu,Nandi:2017rhy}.

A final possibility is to implement EOSs leading to 
mass-radius relations with disconnected branches such as in the twin-stars \citep{Christian:2017jni,Benic:2014jia} or in the two families scenario, in which hadronic and quark stars coexist \citep{Drago:2013fsa,Drago:2015cea,Burgio:2018yix,DePietri:2019khb}. Interestingly, a very recent analysis on the short gamma ray bursts has provided hints in favour of quark stars being the remnants of the mergers of two compact stars \citep{Sarin:2020pwr}.

\acknowledgments
We would like to thank Alessandro Drago and Pietro Bergamini for useful discussions. P. Char also acknowledges the financial support from INFN, as a postdoctoral fellow.


\bibliographystyle{aasjournal}
\bibliography{mybiblio}

\begin{figure}
    \centering
    \includegraphics[height=0.8\textheight]{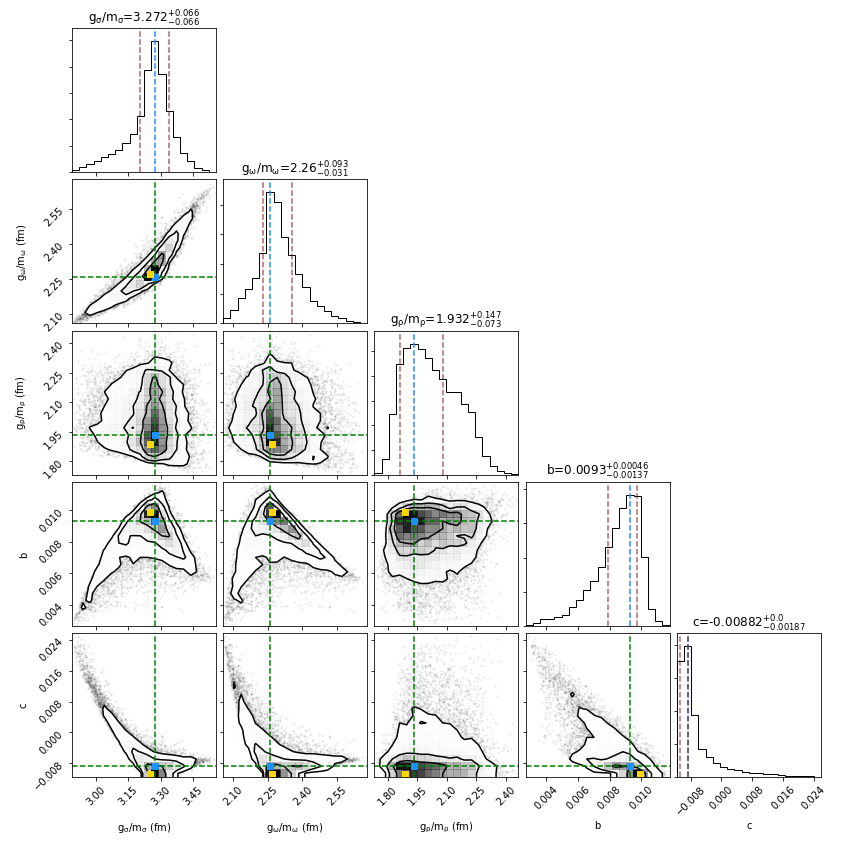}  
    \caption{Distribution of RMF parameters corresponding to the empirical parameters of Fig. \ref{NPP_unif_mode}. {\bf The complete figure set (six images for the other six priors) is available in the online journal.}}
    \label{RMFP_unif_mode}
\end{figure}

\appendix
In this appendix, we provide further calculations for Marg\_unif, Marg\_Gauss priors and the Wide\_unif with $\Lambda$. We also present the results for RMF coupling constants for all the cases: in Table \ref{tab:mode_RMFP}, the most probable values of the joint PDF are listed while  Figure \ref{RMFP_unif_mode} and the corresponding figure set (in the online journal) display their marginalized PDFs.  In Table. \ref{tab:median_NPP}, we provide the median values of the marginalized distributions of the empirical parameters. 

\begin{table}[h!]
    \begin{tabular}{| c | c | c | c | c | c | c |}
    \hline
    \hline 
    Models & $m^{\ast}$ & K (MeV) & $n_{0}$ (fm$^{-3}$) & S (MeV) & $E_0$ (MeV) & L (MeV) \\
    \hline 
    Baseline  & $0.761_{-0.016}^{+0.015}$ & $212.7_{-9.8}^{+32.0}$ & $0.157_{-0.004}^{+0.002}$ & $30.3_{-1.7}^{+2.7}$ & $-16.26_{-0.17}^{+0.17}$ & $84.4_{-5.1}^{+8.3}$ \\
    \hline
    Marg\_unif  & $0.705_{-0.008}^{+0.004}$ & $226.6_{-5.7}^{+13.0}$ & $0.152_{-0.002}^{+0.001}$ & $33.4_{-1.7}^{+3.0}$ & $-16.24_{-0.08}^{+0.08}$ & $97.6_{-5.0}^{+8.8}$ \\
    \hline 
    Marg\_Gauss  & $0.707_{-0.014}^{+0.016}$ & $161.1_{-11.9}^{+14.0}$ & $0.151_{-0.002}^{+0.002}$ & $34.7_{-2.7}^{+2.9}$ & $-16.24_{-0.06}^{+0.06}$ & $101.7_{-8.1}^{+8.6}$\\
    \hline 
    Wide\_unif  & $0.740_{-0.027}^{+0.023}$ & $203.5_{-24.9}^{+82.7}$ & $0.166_{-0.009}^{+0.005}$ & $22.7_{-4.1}^{+11.7}$ & $-15.57_{-1.02}^{+1.17}$ & $62.7_{-12.4}^{+35.3}$ \\
    \hline 
    Wide\_Gauss  & $0.746_{-0.024}^{+0.014}$ & $177.3_{-10.9}^{+18.1}$ & $0.154_{-0.004}^{+0.004}$ & $32.8_{-4.8}^{+4.5}$ & $-16.20_{-0.37}^{+0.37}$ & $93.3_{-14.1}^{+13.6}$  \\
    \hline 
    Wide\_unif with $\Lambda$  & $0.683_{-0.029}^{+0.024}$ & $355.9_{-58.7}^{+44.0}$ & $0.169_{-0.006}^{+0.003}$ & $18.9_{-1.0}^{+2.5}$ & $-15.40_{-1.14}^{+1.09}$ & $54.9_{-4.9}^{+8.7}$  \\
    \hline   
    Baseline with $\Lambda$ & $0.706_{-0.004}^{+0.006}$ & $292.1_{-9.1}^{+5.4}$ & $0.143_{-0.001}^{+0.001}$ & $28.9_{-0.6}^{+1.0}$ & $-16.24_{-0.18}^{+0.17}$ & $83.6_{-2.0}^{+3.0}$  \\
    \hline   
    \end{tabular}
    \caption{Median of the marginalized distributions for the empirical parameters.}
    \label{tab:median_NPP}
\end{table}

\begin{table}[h!]
    \centering
    \begin{tabular}{| c | c | c | c | c | c | c |}
    \hline
    \hline 
    Models & $g_{\sigma}$/$m_{\sigma}$ (fm) & $g_{\omega}$/$m_{\omega}$ (fm) & $g_{\rho}$/$m_{\rho}$ (fm) & b & c \\
    \hline 
    Baseline  & $3.252$ & $2.270$ & $1.888$ & $0.00991$ & $-0.01083$  \\
    \hline
    Marg\_unif  & $3.505$ & $2.629$ & $2.047$ & $0.00548$ & $-0.00615$  \\
    \hline 
    Marg\_Gauss  & $3.585$ & $2.628$ & $2.267$ & $0.00713$ & $-0.00985$ \\
    \hline 
    Wide\_unif  & $3.351$ & $2.296$ & $1.774$ & $0.01148$ & $-0.01510$  \\
                & $3.000$ & $2.153$ & $0.882$ & $0.00596$ & $0.00231$  \\
    \hline 
    Wide\_Gauss  & $3.381$ & $2.351$ & $2.490$ & $0.01056$ & $-0.01373$   \\
    \hline 
    Wide\_unif with $\Lambda$  & $3.340$ & $2.609$ & $0.274$ & $0.00279$ & $-0.00129$   \\
    \hline    
    Baseline with $\Lambda$  & $3.556$ & $2.768$ & $1.971$ & $0.00319$ & $-0.00180$   \\
    \hline  
    \end{tabular}
    \caption{Most probable RMF coupling constants from the joint PDF.}
    \label{tab:mode_RMFP}
\end{table}

\begin{figure}
    \centering
    \includegraphics[height=0.8\textheight]{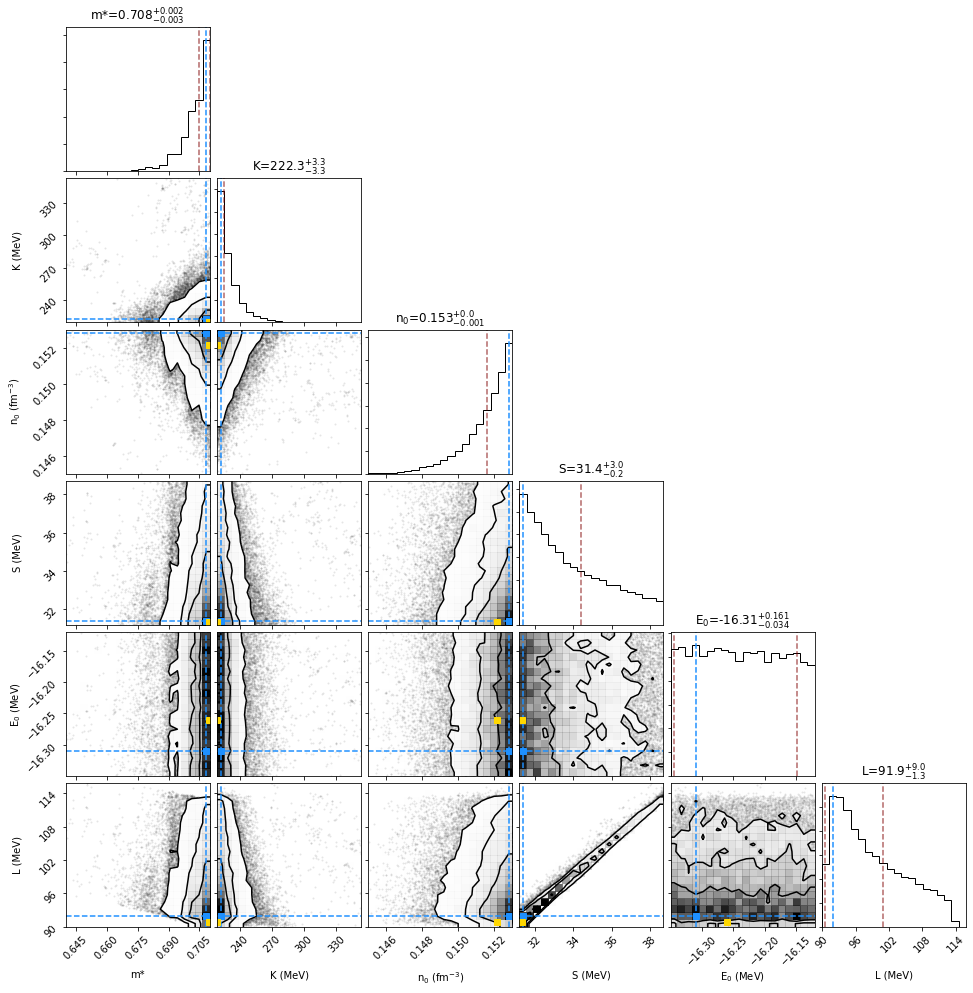}   
    \caption{Same as Figure \ref{NPP_unif_mode} for the Marg\_unif prior}
    \label{NPP_marg_unif_mode}
\end{figure}

\begin{figure}
    \centering
        \begin{tabular}{cc}
         \includegraphics[width=0.5\textwidth]{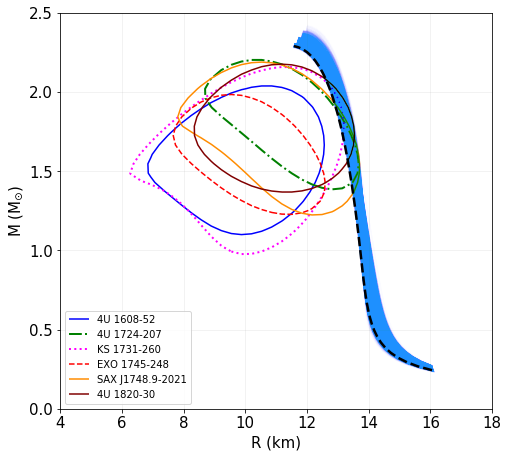}& \includegraphics[width=0.5\textwidth]{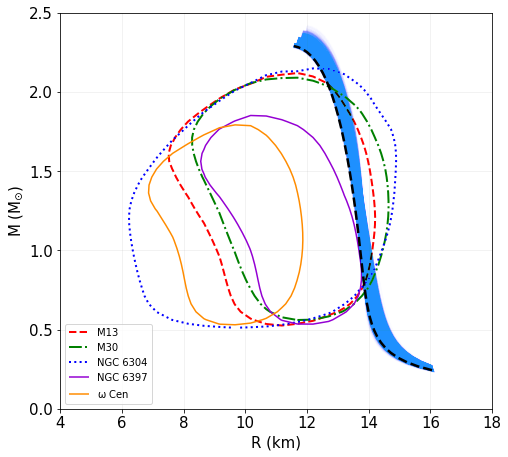} \\
        \includegraphics[width=0.5\textwidth]{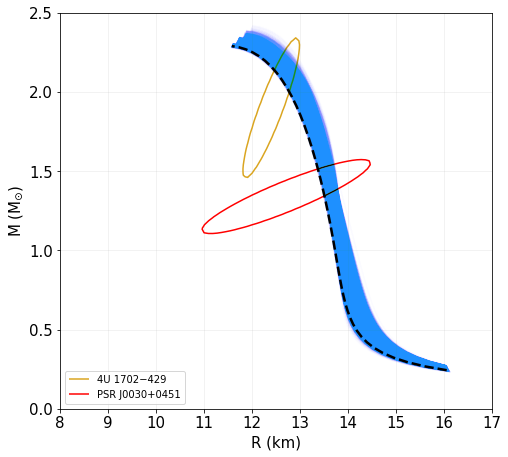} & 
        \includegraphics[width=0.5\textwidth]{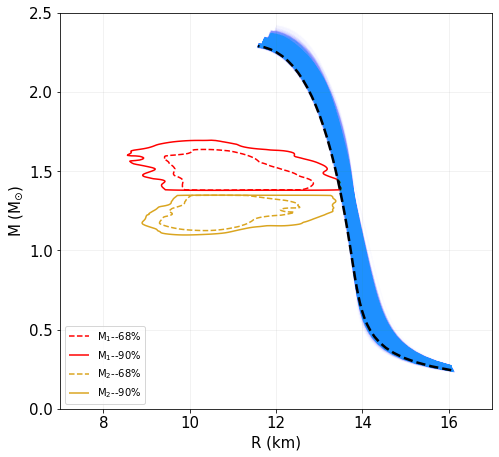}\\
    \end{tabular}
        \caption{Same as Figure \ref{fig:unif_mr} for the Marg\_unif prior.}
    \label{fig:marg_unif_mr}
\end{figure}

In Figure \ref{NPP_marg_unif_mode}, we show the results for Marg\_unif prior. Within this particular range, one can usually construct EOSs that are stiffer with respect to the ones obtained with the  baseline prior. But, to conform with the observational data, the most probable parameter set for this prior again sits on the edge of the boundary where one can get the softest possible EOS. Compared to the baseline, the difference we get is in the marginalized distribution of $m^*$ which does not have a peak for Marg\_unif case. Rather, it shows the trend of having the most probable configuration near the edge of the upper limit corresponding to the softest EOS. Otherwise, the trends for  $K$, $n_0$, and $S$ are similar to those of our baseline prior. In Fig. \ref{fig:marg_unif_mr}, the inferred M-R sequences are shown along with the most probable one. Due to our choice of the parameter ranges, the inferred EOSs are stiffer and the M-R sequences are outside of the 68\% regions of most of the X-ray sources and the GW data. 

\begin{figure}
    \centering
    \includegraphics[height=0.8\textheight]{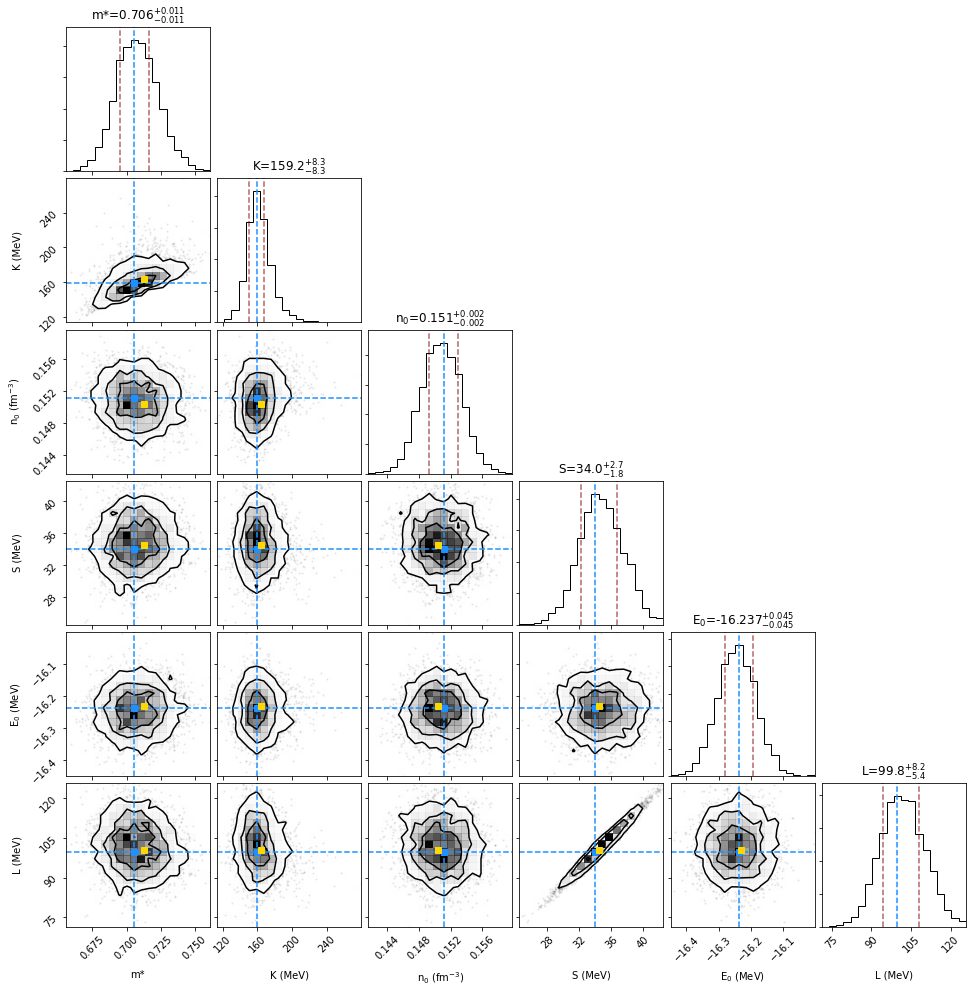}   
    \caption{Same as Figure \ref{NPP_unif_mode} for the Marg\_Gauss prior. }
    \label{NPP_marg_gau_mode}
\end{figure}

\begin{figure}
    \centering
        \begin{tabular}{cc}
         \includegraphics[width=0.5\textwidth]{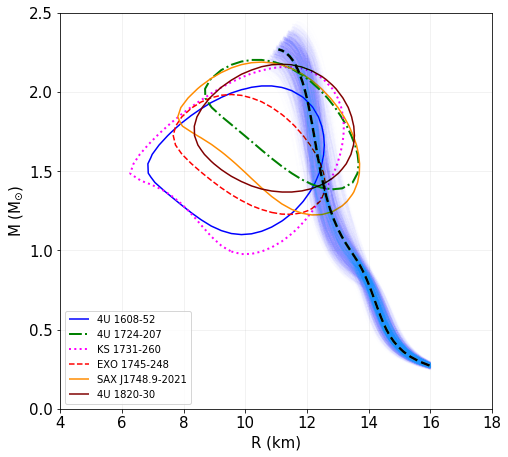}& \includegraphics[width=0.5\textwidth]{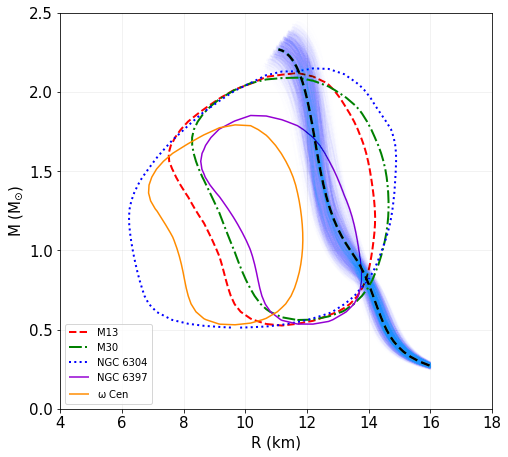} \\
        \includegraphics[width=0.5\textwidth]{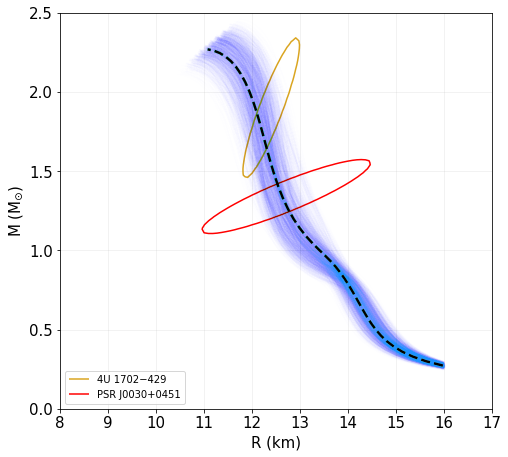} & 
        \includegraphics[width=0.5\textwidth]{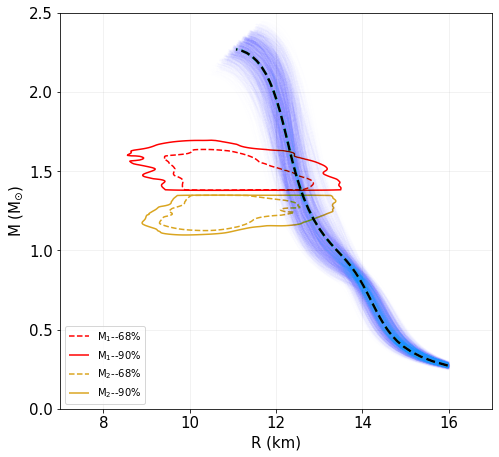}\\
    \end{tabular}
        \caption{Same as Figure \ref{fig:unif_mr} for the Marg\_Gauss prior.}
    \label{fig:marg_gau_mr}
\end{figure}

In Figure \ref{NPP_marg_gau_mode}, the distribution of the empirical parameters are shown for the Marg\_Gauss prior. The peak of $K$ is shifted by more than $3\sigma$ from the the peak of the prior. The corresponding mass-radius relations shown in Figure \ref{fig:marg_gau_mr} are qualitatively similar to Wide\_Gauss prior. But, the maximum mass for the most probable configuration is much higher, as the EOS becomes stiffer due to fact that the value of the $m^*$ is peaked at a much lower value.

\begin{figure}
    \centering
    \includegraphics[height=0.8\textheight]{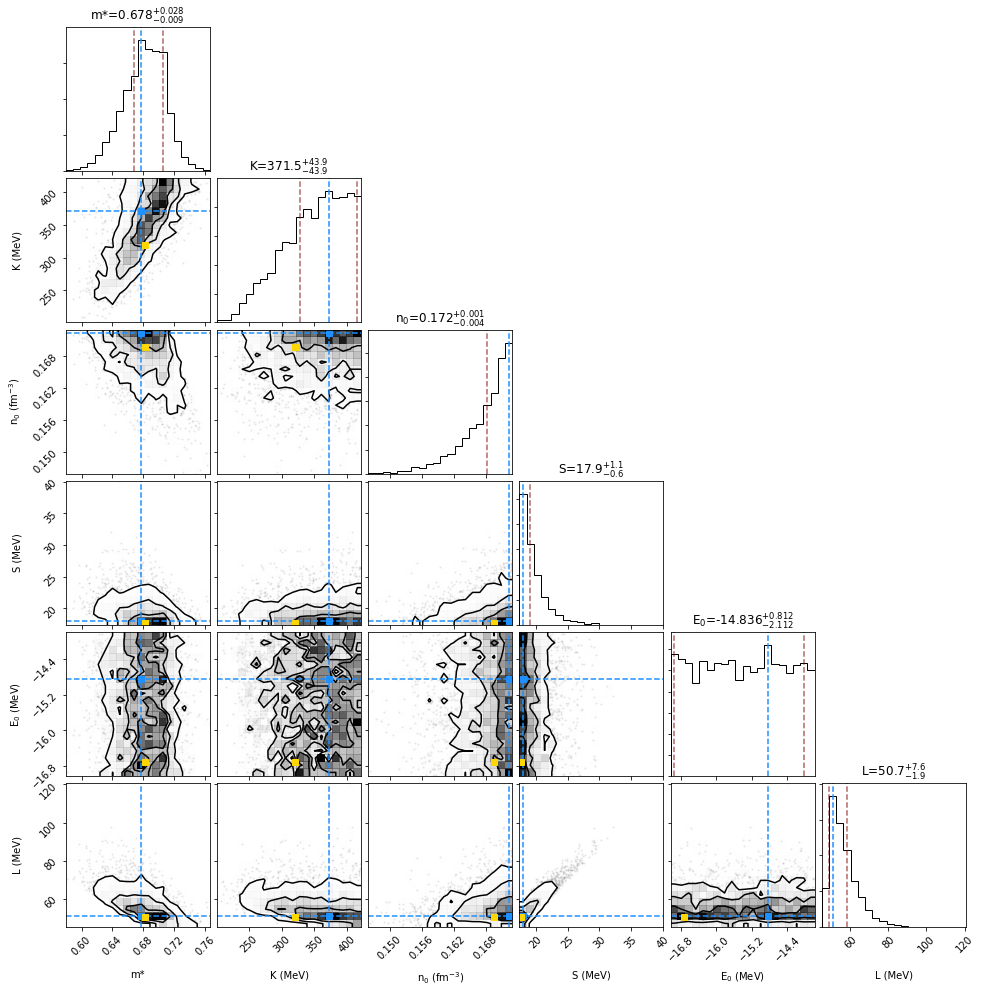}   
    \caption{Same as Figure \ref{NPP_unif_mode} for the Wide\_unif with $\Lambda$ in the system}
    \label{NPP_lam_mode}
\end{figure}

\begin{figure}
    \centering
        \begin{tabular}{cc}
         \includegraphics[width=0.5\textwidth]{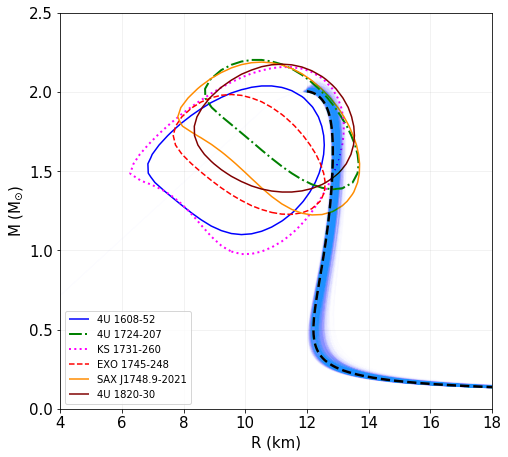}& \includegraphics[width=0.5\textwidth]{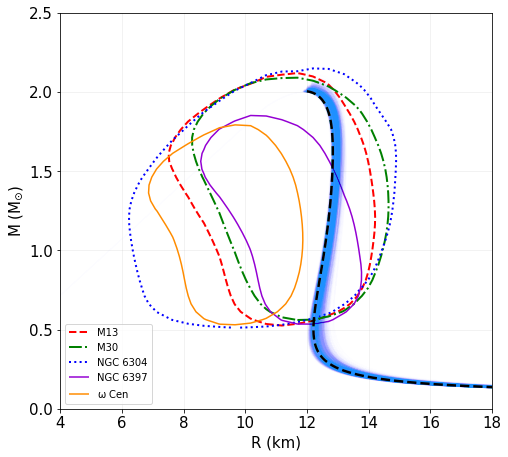} \\
        \includegraphics[width=0.5\textwidth]{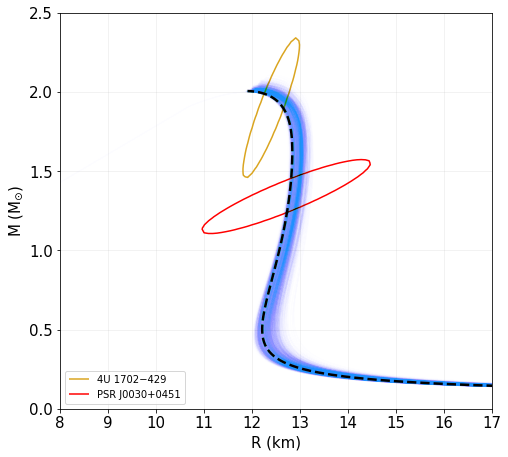} & 
        \includegraphics[width=0.5\textwidth]{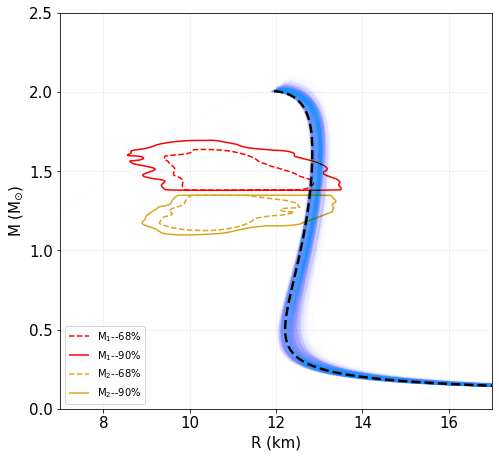}\\
    \end{tabular}
        \caption{Same as Figure \ref{fig:unif_mr} for the Wide\_unif with $\Lambda$ in the system}
    \label{fig:lam_mr}
\end{figure}

The posterior distributions for the Wide\_unif with $\Lambda$ case are shown in Figure \ref{NPP_lam_mode}, and the mass radius sequences in Figure \ref{fig:lam_mr}. The preferred $K$ value is very high to reduce the softening of the EOS due to appearance of hyperons and the $S$ is very small to make the matter more symmetric i.e. less neutron rich which in turn suppresses the production of $\Lambda$'s. But, such small values of $S$ are ruled out by the present experimental knowledge.

    \figsetstart



    \figsetgrpstart

    \figsetgrpnum{figurenumber.1}

    \figsetgrptitle{short image 1 title}

    \figsetplot{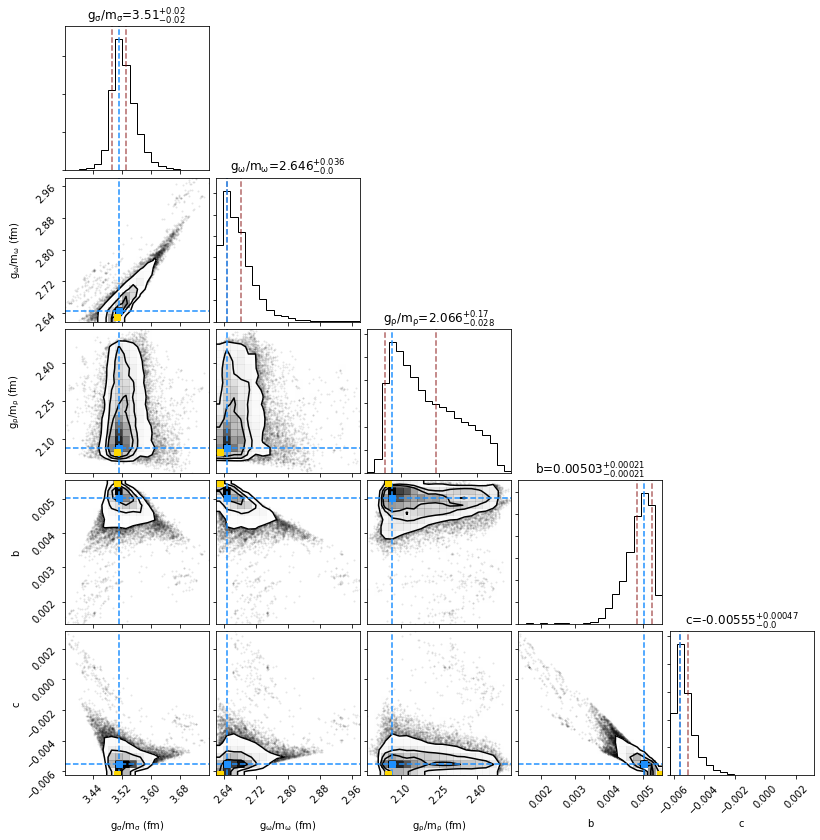}

    \figsetgrpnote{Same as Figure \ref{RMFP_unif_mode} corresponding to the empirical parameters of Fig. \ref{NPP_marg_unif_mode}}

    \figsetgrpend

    \figsetgrpstart

    \figsetgrpnum{figurenumber.2}

    \figsetgrptitle{short image 2 title}

    \figsetplot{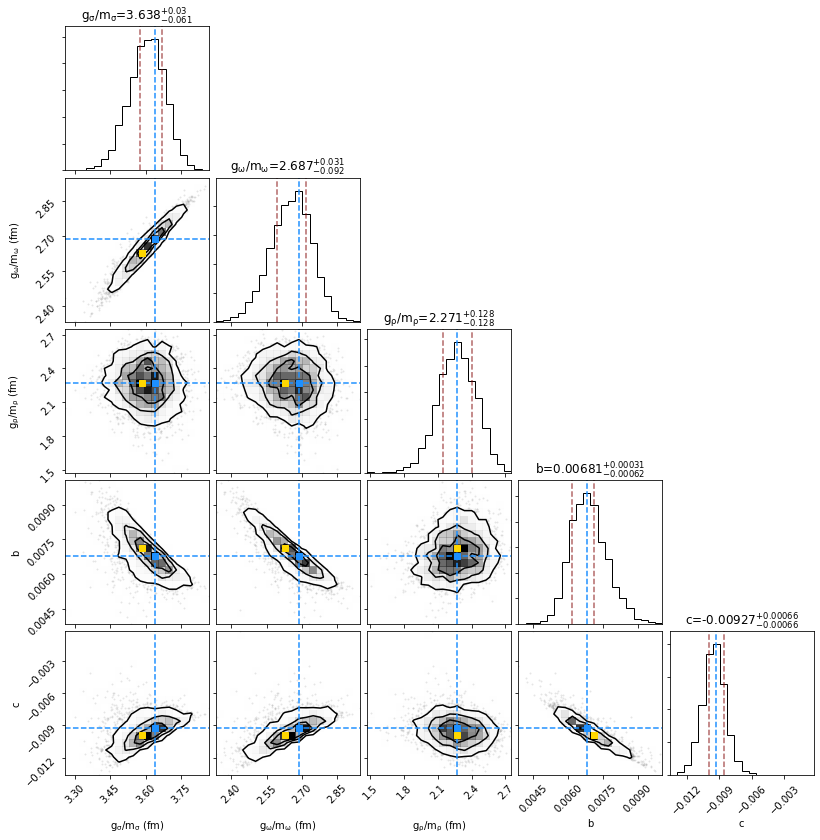}

    \figsetgrpnote{Same as Figure \ref{RMFP_unif_mode} corresponding to the empirical parameters of Fig. \ref{NPP_marg_gau_mode}}

    \figsetgrpend
    
    \figsetgrpstart

    \figsetgrpnum{figurenumber.3}

    \figsetgrptitle{short image 3 title}

    \figsetplot{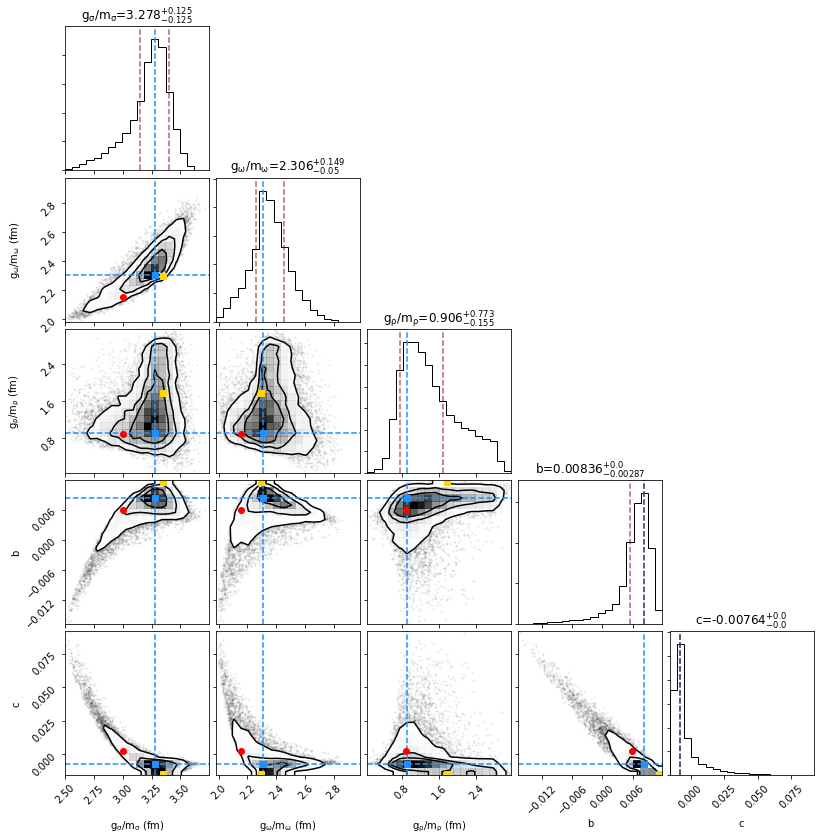}

    \figsetgrpnote{Same as Figure \ref{RMFP_unif_mode} corresponding to the empirical parameters of Fig. \ref{NPP_unif_wide_mode}}

    \figsetgrpend

    \figsetgrpstart

    \figsetgrpnum{figurenumber.4}

    \figsetgrptitle{short image 4 title}

    \figsetplot{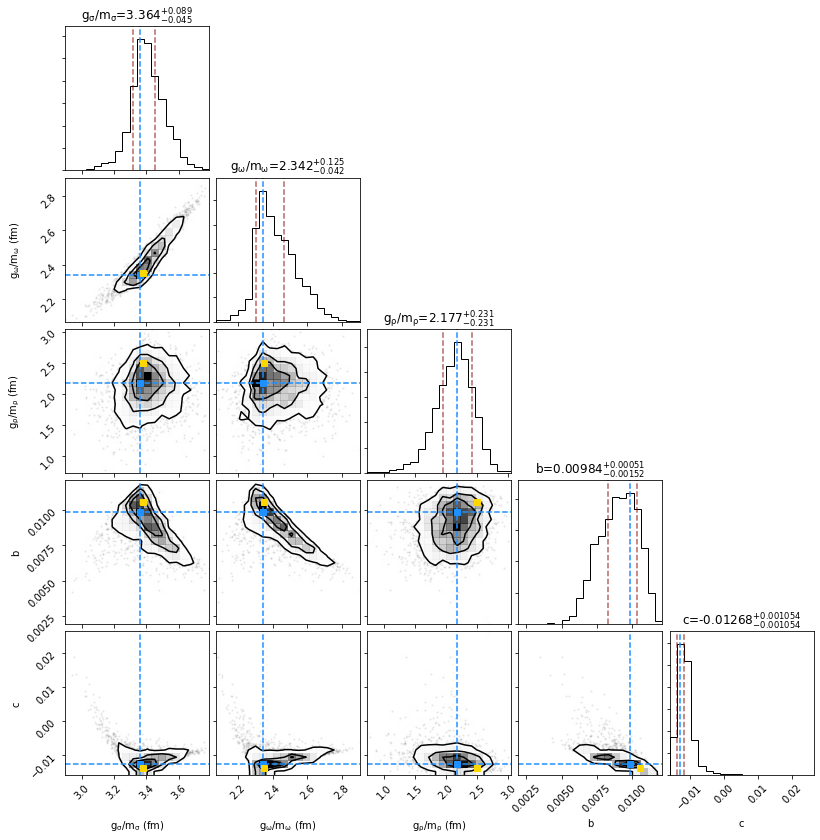}

    \figsetgrpnote{Same as Figure \ref{RMFP_unif_mode} corresponding to the empirical parameters of Fig. \ref{NPP_gau_wide_mode}}

    \figsetgrpend

    \figsetgrpstart

    \figsetgrpnum{figurenumber.5}

    \figsetgrptitle{short image 5 title}

    \figsetplot{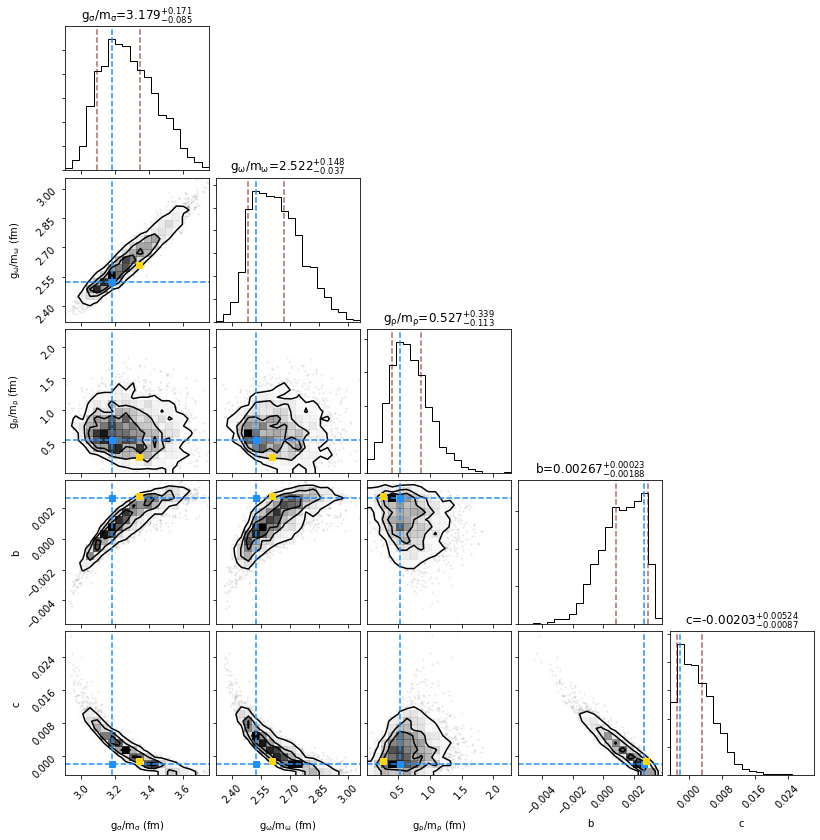}

    \figsetgrpnote{Same as Figure \ref{RMFP_unif_mode} corresponding to the empirical parameters of Fig. \ref{NPP_lam_mode}}

    \figsetgrpend

    \figsetgrpstart

    \figsetgrpnum{figurenumber.6}

    \figsetgrptitle{short image 6 title}

    \figsetplot{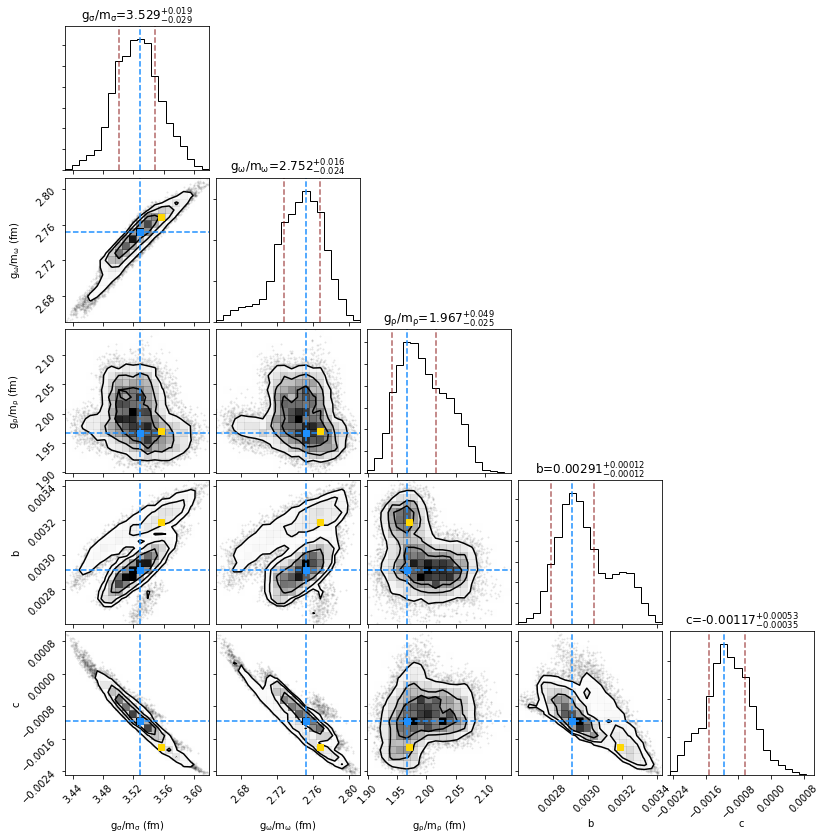}

    \figsetgrpnote{Same as Figure \ref{RMFP_unif_mode} corresponding to the empirical parameters of Fig. \ref{NPP_lam_glen_mode}}

    \figsetgrpend
    
    \figsetend

\begin{figure}[!ht]
	\begin{centering}
		\epsfig{file=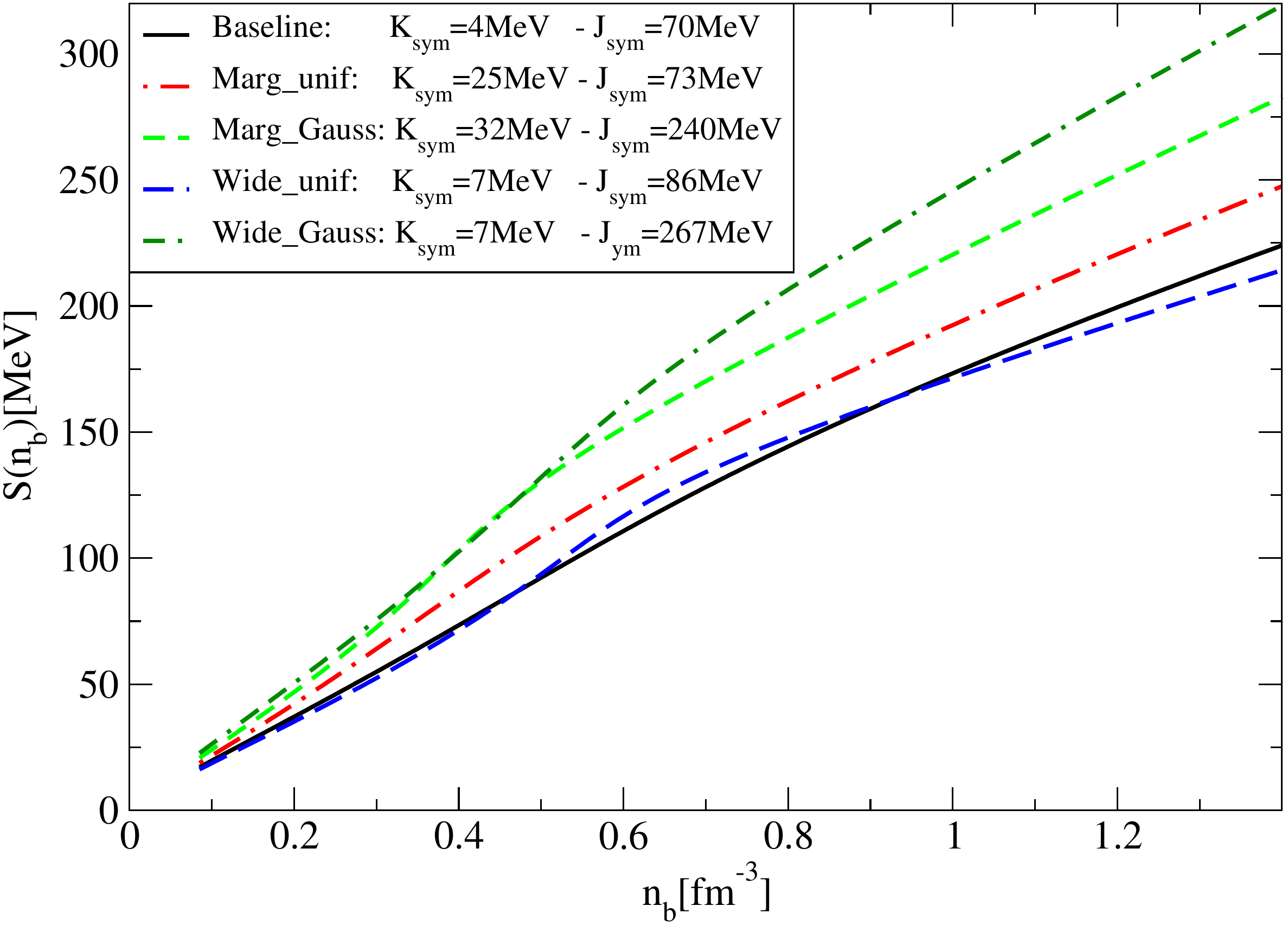,height=7cm,width=11cm}
		\caption{Density dependence of symmetry energy for the same EOS of Figure \ref{masseff}. We indicate also the high order derivatives $K_{\text{sym}}$ and $J_{\text{sym}}$. }
		\label{neutronmatter}
	\end{centering}
\end{figure}

\end{document}